\documentclass{aastex}
\usepackage{emulateapj5}

\def\m{\mbox {MSDM}}

\def\h1{\mbox {\rm HI}}

\def\fig{{Figure}}

\def\deg{\mbox {$^{\circ}$}~}
\def\msun{\mbox {${\rm ~M_\odot}$}}

\def\msunyr{\mbox {$~{\rm M_\odot}$~yr$^{-1}$}}
\def\msunyrk2{\mbox {$~{\rm M_\odot}$~yr$^{-1}$~kpc$^{-2}$}}
\def\msunpc2{\mbox {${\rm ~M_\odot ~pc}^{-2}$}}
\def\angs{\mbox {~\AA}}
\def\lya{\mbox {Ly$\alpha$~}}
\def\Ha{\mbox {H$\alpha$~}}
\def\Hb{\mbox {H$\beta$~}}
\def\Hg{\mbox {H$\gamma$~}}

\def\asec{\ifmmode {'' }\else $''~$\fi}  
\def\amin{\ifmmode {' }\else $'~$\fi}    
\def\sles{\lower2pt\hbox{$\buildrel {\scriptstyle <}
   \over {\scriptstyle\sim}$}} 
\def\sgreat{\lower2pt\hbox{$\buildrel {\scriptstyle >}
    \over {\scriptstyle\sim}$}} 

\def\kms{\mbox {~km~s$^{-1}$} }
\def\ergsec{~ergs~s$^{-1}$~}

\def\flux{~erg~s$^{-1}$~cm$^{-2}$}
\def\cm3{~cm$^{-3}$}
\def\cm5{~cm$^{-5}$}

\def\um{\mbox {$\mu${\rm m}} }

\def\et{{\rm et\thinspace al.}\ }   

\def\apj{ApJ}
\def\apjs{ApJS}
\def\pasp{PASP}
\def\aj{AJ}
\def\mn{MNRAS}

\def\aa{A\&A}

\slugcomment{To appear in ApJ.}
\shorttitle{Spectroscopic Search for LAEs}
\shortauthors{MSDM}

\begin{document}



\title{A MAGELLAN IMACS SPECTROSCOPIC SEARCH FOR \lya EMITTING GALAXIES
AT REDSHIFT 5.7}

\author{Crystal L. Martin\altaffilmark{1}}
\affil{University of California, Santa Barbara}
\affil{Department of Physics}
\affil{Santa Barbara, CA, 93106}
\email{cmartin@physics.ucsb.edu}

\and
\author{Marcin Sawicki}
\affil{St. Marys University}
\affil{Department of Astronomy and Physics}
\affil{923 Robie Street}
\affil{Halifax, N.S., B3H 3C3, Canada}

\and
\author{Alan Dressler and
Pat McCarthy}
\affil{OCIW, Santa Barbara Street, Pasadena, CA}


\altaffiltext{1}{Packard Fellow}


\begin{abstract}
We present results from a blind, spectroscopic survey for $z \sim 5.7$
\lya -emitting galaxies using the  Inamori Magellan Areal Camera and
Spectrograph.  A total of  $\sim 200$  square arcminutes were observed
in the COSMOS and LCIRS fields using a narrowband filter, which transmits
between atmospheric emission lines at 8190\AA, and a mask with 100 longslits.
This observing technique provides higher emission-line sensitivity
than narrowband imaging and probes larger volumes than strong lensing.
We find 170 emission-line galaxies and identify their redshifts
spectroscopically. We confirm three \lya\ emitting galaxies (LAEs),
the first  discovered using multislit-narrowband spectroscopy. Their line
profiles are narrow, but fitted models suggest instrinsic, unattenuated
widths $\sim 400$\kms FWHM. The red wing of the line profiles present features
consistent with galactic winds. The star formation rates of these galaxies
are at least 5-7\msunyr\ and likely a factor of two higher.
We estimate the number density of $L \ge  5 \times 10^{42}$\ergsec\ LAEs
is $9.0^{+12}_{-4} \times 10^{-5} h_{70}^3$~Mpc$^{-3}$ at redshift 5.7
and constrain the
Schechter function parameters describing this population.  Galaxies
fainter than our detection limit may well be the primary source of
ionizing photons at $z \sim 6$.  We argue, however, that the break
luminosity $L_{*,Ly\alpha}$ is not yet well constrained.
If this break luminosity is near our detection limit, and somewhat lower
than previous estimates, then the detected LAE population could be
responsible for ionizing the intergalactic gas at redshift $z \sim 6$.
We discuss the potential of multislit-narrowband spectroscopy for
deeper emission-line surveys.
\end{abstract}



\section{INTRODUCTION}

Reionization of the intergalactic medium (IGM) marks a phase transition in
the baryonic component of the universe and quite likely heralds the 
emergence of the first galaxies that sustain star formation beyond a 
single burst (Oh \& Haiman 2001). The appearance of complete \lya\ 
absorption troughs in quasar spectra at $z \sgreat\ 6$ shows that 
at least 1 in $10^4$ hydrogen atoms was still neutral then
(Becker \et 2001; Djorgovski \et 2001; Fan \et 2002; 
Songaila 2004; Fan \et 2006), although much of the IGM appears to have been ionized 
by $z \sim 6.5$ (Becker \et 2006, 2007). The Thomson optical depth
measured from the microwave background polarization suggests an
earlier start to reionization at $z_r = 11^{+3}_{-3}$ (Page \et 2006; 
Spergel \et 2007).  The \lya\ emission line, surprisingly, is proving
very useful for  identifying galaxies over this protracted period of 
Reionization (Iye et al. 2006; Stark \et 2007b). Frequency shifting across 
the Stromgren spheres of the sources allows a fraction of the emission line 
photons to propagate through an IGM with a considerable neutral fraction.
Understanding the intrinsic evolution of \lya emitters (LAEs) builds a
foundation for using the apparent evolution of the LAE luminosity function 
during reionization to measure the changing neutral fraction 
(Malhotra \& Rhoads 2004; Kashikawa \et 2006).

High-redshift LAEs may provide the first glimpse of primeval 
galaxies, owing to the high \lya\ equivalent-width of young, metal-poor 
galaxies (Malhotra \& Rhoads 2002; Dawson \et 2004; Schaerer 2007).
The stellar mass density at $z \sim 5$ (Stark \et 2007a), 
the presence of carbon in the IGM at $z = 5.7$ (Simcoe 2006; Ryan-Weber,
Pettini, \& Madau 2006), and the steep faint-end slope of the i-dropouts 
(Bouwens \et 2007)
can all be interpreted as evidence for a 
significant population of as yet undetected dwarf galaxies at high
redshift. The properties of these low-mass galaxies during and just after 
reionization are cosmologically important. Finding them and measuring their 
properties will continue to be of interest well into the next decade.



Finding high-redshift galaxies remains largely a contrast problem
with the sky.  Selection by \lya\ emission provides higher sensitivity
than continuum break identification for populations with large line
equivalent widths, like young, metal-poor galaxies.
Over the past decade, several tens of LAEs have been confirmed 
near redshift 6 among objects found via narrowband 
imaging (Hu \et 2002; Cuby \et 2003; Ajiki \et 2003; 
Kodaira \et 2003; Hu \et 2004; Rhoads \et 2004; Ajiki \et 2004; 
Taniguchi \et 2005; Shimasaku \et 2006; Kashikawa \et 2006; Murayama \et 2007; 
Westra \et 2005, 2006; Ouchi \et 2007). The redshifts of nearly all of these 
galaxies shift the \lya line into one of the bandgaps in the OH (hydroxyl)
spectrum of the terrestrial sky.  The dark regions at 8200\AA\ and 9200\AA\
are the widest ($\sim 150\AA$) and are useful for studying the end
of reionization. Lyman-alpha emission at $z =  5.7$ and 6.5
redshifts into these bands.

Spectroscopic surveys also benefit from these low background regions and
offer higher sensitivity than narrowband imaging. Dispersion of the background 
improves line-to-sky contrast, although a smaller solid angle of sky is covered.
One prolific spectroscopic approach has been targetting strongly lensed LAEs.
Probing comparatively tiny volumes, three very faint LAEs have been confirmed 
with this technique at $z \sim 5-6$ (Ellis \et 2001; Santos \et 2004). 
Using the Hubble Space Telescope to get above the atmospheric OH emission
has also produced 3 LAEs at $z > 5$ (Pirzkal \et 2007). Serendipitous
discoveries have also been made in the ``sky region'' of longslit 
spectra (Sawicki \et 2007; Stern \et 2007). In this paper, we
present the first detections of LAEs with a technique --
multislit-narrowband spectroscopy (MNS) -- designed to optimize 
the sensitivity of ground-based telescopes to high-redshift LAEs.
Previous surveys with MNS failed to find any LAEs (e.g. Crampton \& Lilly 1999; 
Martin \& Sawicki 2004; Tran et al. 2004).

Observations were made with the  Inamori Magellan Areal Camera and Spectrograph 
(IMACS, Dressler \et 2006) on the Baade Telescope, which offered a wide field of
view.    Longslit spectra were obtained in the dark region at 8200\AA.  A
custom blocking filter and ``Venetian blinds'' mask multiplexed the effective
area one-hundred fold.  We carried out
four blind searches: two in the COSMOS (Cosmic Evolution Survey) field 
and two in a LCIRS (Las Campanas Infrared Survey) field. 
The total survey area of 200 square arcminutes is significantly 
larger than that attempted previously with similar techniques.

We describe the survey in \S2 and explain how we found 
emission-line objects and identified their redshifts in \S3.  
Some readers may wish to skip directly to \S4, where we discuss the star 
formation and feedback properties of the LAEs, estimate their space 
density, and explore the contribution of the LAE population to the ionization 
of the IGM at $z =5.7$. We summarize the advantages and disadvantages
of MNS in \S5, including a direct comparison to planned instrumentation
for the next generation of telescopes.  Section~\ref{sec:summary}
summarizes our conclusions. Properties of the 
foreground line-emitters will be discussed in a separate paper.
We adopt a cosmology with $\Omega_m = 0.3$, $\Omega_{\Lambda} = 0.7$, 
$H_0=70$~km~s$^{-1}$~Mpc$^{-1}$, and $\Omega_b h^2 = 0.02$
throughout this paper. The angular scale at redshift 5.7 is 
then 5.87~kpc/\arcsec ;  the universe is 0.98~Gyr
old; and the time between redshift 6.5 and 5.7 is 151~Myr.

\section{OBSERVATIONS AND DATA ANALYSIS} \label{sec:observations}

Four types of data were used in our emission-line survey. We began
with the blind, emission-line survey with the IMACS instrument 
on the Baade telescope. Broad bandpass spectroscpy of emission-line
objects was then obtained to identify the redshift of the line via
the presence (or absence) of other lines. A subsample of LAE candidates 
was then followed up with higher-resolution spectroscopy to 
determine the shape of the line profile.  Broad-band imaging of
our two fields by the COSMOS (Scoville \et 2007; Taniguchi 
\et 2007) and the LCIRS (Marzke \et 1999; Chen \et 2002) supplement
our new spectroscopic observations.  The photometric redshifts from COSMOS 
(Mobasher et al. 2007) were particularly useful
because the association of a line-selected object with a
continuum source is often ambiguous.

\subsection{Multislit, Narrowband Spectroscopic Search} \label{sec:search}

\begin{figure}[h]
      {\includegraphics[scale=0.4,angle=-90,clip=true]{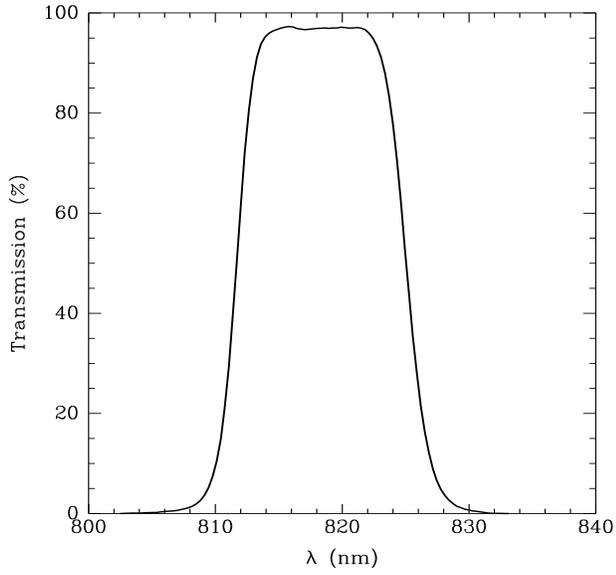} 
         }
          \caption{Transmission of the NB8190 filter.
          }
           \label{fig:filter} \end{figure}

\begin{figure}[h]
      {        \includegraphics[scale=0.4,angle=-90,clip=true]{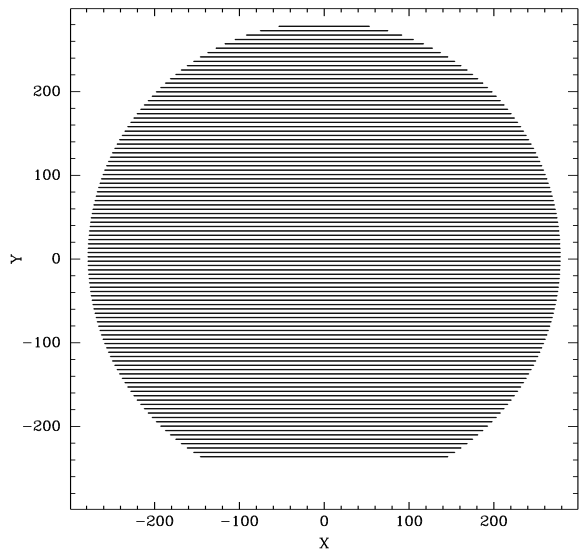} 
         }
          \caption{IMACS slit mask used for the spectroscopic survey.
	    Coordinates denote positions in the focal plane. 
	    The 1\farcs5 wide slits subtend a geometric area of 
	    55.3 $\square ^{\arcmin}$ per pointing within the 
	    27\farcm50 diameter field of the camera.
          }
           \label{fig:mask} \end{figure}

Multislit narrowband spectroscopy remains a relatively new technique,
and we illustrate the filter transmission in Figures~\ref{fig:filter}
and the mask design in Figure~\ref{fig:mask}.
Our custom blocking filter fabricated by Barr Associates passes
a 121\angs\ FWHM bandpass at 8188\angs. The transmission is
over 90\% across this band and drops sharply to less than 
5\% transmission outside the $\lambda 8104$ -- $\lambda 8273$
bandpass. We placed this filter in the parallel beam directly
in front of the 200~lines mm$^{-1}$ grism and recorded data
with the short, f/2, camera. With a dispersion of  2.0 \angs ~pix$^{-1}$,
we fit spectra from 100 longslits (numbered 9 to 108) on the
eight 2k by 4k thinned, SITe detectors.
         At low dispersion, the angle between the first-order
	 spectrum and the zero-order image of the mask is smaller than
	 the detector,  so a zero-order image of the mask also falls
         on the detector. This zero-order image is easily recognized 
	 via the pattern of the multislit mask, and this region is 
	 masked and ignored in the analysis.
The 1\farcs5 wide slits subtend a geometric area of 55.3 $\square ^{\arcmin}$ per 
pointing within the 27\farcm50 diameter field of the camera.
With this setup, emission lines from diffuse 
sources have a width of $12.8 \pm 0.3$\angs\ FWHM. The spectral resolution
is typically a factor of two better for spatially-unresolved objects,
as determined by the atmospheric seeing.

Table~\ref{tab:observations} summarizes the search observations.
Two masks were exposed in each of two fields.  The masks were
offset by half the slit separation. Exposure times ranged from
6 to 11 hours, and the image quality was generally $0\farcs8$
FWHM or better. Image quality was good across the detector during the 2005 
March and May runs, which used the field corrector and atmospheric dispersion 
compensator.  The 2004 April run took place prior to installation of the
ADC/corrector, and the image quality was further compromised by an un-even 
focus across the detector caused by a misaligned tertiary mirror.

We used techniques standard to optical {\it imaging} to remove 
the instrumental
signature from the data. The bandpass is narrow enough to ignore the wavelength 
dependence of the detector response; and  we corrected for pixel-to-pixel 
response variations with direct images of the flatfield screen taken through  
the NB8190 filter (in the filter wheel). Stacked images were constructed using 
cosmic ray masks. Faint, sky lines are present even in the relatively clean, 
dark bandpass at 8200\AA. We subtracted this sky background after computing 
it in a local aperture along the slit direction. Subtraction of this sky 
model revealed the 
objects. These procedures were implemented with a set of custom scripts, 
implemented with PerlDL\footnote{The
      Perl Data Language (PDL) has been developed by K. Glazebrook, J.
      Brinchmann, J. Cerney, C. DeForest, D. Hunt, T. Jenness, T. Luka, R.
      Schwebel, and C. Soeller and provides a high-level numerical
      functionality in the perl scripting language (Glazebrook \& Economou,
      1997). It can be obtained from http:$\backslash \backslash$
      pdl.perl.org.}
 and IRAF routines.\footnote{
     The Image Reduction and Analysis Facility, a general purpose 
     software system for the reduction and analysis of astronomical data. 
     IRAF is written and supported by the National Optical Astronomy 
     Observatories, which is operated by the  Association of 
     Universities for Research in Astronomy, Inc. under cooperative 
     agreement with the National Science Foundation.}

We used the location of each object along the slit and the transformation 
from mask to sky coordinates, kindly provided by K. Clardy, to determine 
object positions in the spatial direction. We tested the accuracy of the
{\it relative astrometry} using bright, foreground objects that fell on slits 
and the broadband images described below.  The position errors along
the slit were normally 
distributed with standard deviation of 0\farcs24 for the 2004 data, 
slightly less  for the 2005 observations.  The same bright objects 
present offsets perpendicular to the slit of $\pm\ 0\farcs36$ ($1\sigma$ result). 
We emphasize, however, the degeneracy between the position of an object
in the slit and the wavelength of the emission line that we measure.
Half a slit width is $\pm\ 0\farcs75$, and offsets approaching this limit
lead to wavelength errors  $\le 7.5\AA$. The {\it accuracy of the zero-point} 
of the astrometric transformation was estimated by comparing the coordinates of 
the mask alignment stars, taken from the USNO catalog, to their coordinates in 
the Subaru images of the COSMOS field. After rejection of two stars,
we find a systematic offset of 0\farcs40 W and and 0\farcs29 S of the 
Subaru positions using four stars.


We inspected the stacked, sky-subtracted frames from each of the four 
narrowband, multislit observations for emission lines.  Line detection was 
then automated with SExtractor (version 2.3.2, Bertin \& Arnouts 
1996), requiring detection over at least 9-contiguous pixels, corresponding to 
a spatial width $ \ge\ 0\farcs6$ and a line width $\ge\  6.4\angs$.  The
SExtractor software recovered our initial, by-eye list of sources; additional
candidates were inspected and found to be associated with incomplete masking
of chip edges, zero-order images, and cosmic-rays.  The SExtractor software
worked well for finding the emission-line spectra because the signal from
a line is very similar to that from an object in a direct image. The
photometry from SExtractor is not appropriate for our survey
however, because sky and continuum subtraction needed to be handled with
techniques specific to spectroscopic data.

We measured the flux of each emission line in a fixed apertures using 
IRAF.  This line photometry was performed on the images prior to sky subtraction 
using apertures defined on the sky-subtracted frames. Because faint,
sky lines are present even in this relatively clean window, care had to
be taken to measure the average sky counts from exactly the same bandpass 
as the line aperture. Continuum was measured in the same spatial aperture
as the line but was usually undetected.  A flux calibration
was derived from observations of spectrophotometric standards taken 
during each run. The throughput during the April 2004 run was 
                28\% better than that during the spring 2005 runs due to a 
		recent re-aluminization of the telescope mirror.
The uncertainty in this calibration is 5-10\% (except for the 2005 May 
search), and our photometric errors are normally dominated by the random 
pixel-to-pixel variations in the background under the object and/or the
Poisson noise from the total counts. The May run was generally clear but
not photometric, and we adopt the flux calibration derived 2 months 
earlier with the identical instrumental setup.

\subsection{Broad Bandpass Spectroscopy} \label{sec:f2spectra}

We obtained spectra of emission-line candidates with 
IMACS, a low resolution grism (usually 200 mm$^{-1}$), the GG455 blocking filter,
and multislit masks with 1\farcs5 slits. Masks for the 10H
field were observed in 2005 May (4.5 hrs) and 2007 March (5.0 hrs).
Masks in the 15H field were observed in 2004 July (7.0 hrs), 
2005 June (2.0 hrs), and 2006 June (3.08 hrs with the  150 mm$^{-1}$
grism).  The wavelength coverage for a typical source was 5500\angs\ to 9500\angs.
The COSMOS\footnote{
        The Carnegie Observatories System for MultiObject Spectroscopy
	was created by A. Oemler, K. Clardy, D. Kelson, and G. Walth; 
	it is available at  http://www.ociw.edu/Code/cosmos .}
data reduction package was used to stack frames, remove
cosmic rays, subtract sky, calibrate wavelength, and 
extract spectra. The 2007 March and 2004 July observations 
of the 10H and 15H fields, respectively, are significantly more 
sensitive than the other follow-up observations.  

To date, low-resolution spectra have been attempted for
95\% of the emission-line objects in both the 10H
and 15H fields with a recovery rate of 80\% and 95\%,
respectively.  Relative astrometric errors 
were of particular concern for pre-corrector discoveries in the 
2004 data, but we found no tendency for unrecovered objects
to be on the periphery of the field.

\subsection{Higher Resolution Spectroscopy}

We obtained higher resolution spectra of \lya candidates
in the 10H Field with the IMACS long, f/4  camera and the
1200 mm$^{-1}$ grating in 2006 February. Line emission
from sources filling the 1\farcs5 slits has a FWHM of 
94\kms;  and the resolution for unresolved objects is 
50 to 63\kms. The two masks included only a subsample of the candidates 
passing the f/2 follow-up. Exposure times of 8.24 hrs and 8.75 hrs
were sufficient under clear conditions to detect most targets.
Some data for the 15H Field were obtained with the same
configuration in  2006 February (6.0 hrs and some cloud
cover) and 2006 June (3.5 hrs with cirrus; and 5.0 hrs 
under clear skies).

Resolved spectra in the 15H field were obtained at Keck in 2006 May using
the 831 mm$^{-1}$ grating on LRIS (Oke \et 1995). Under thin clouds,
integration times of 1.5 to 3 hrs yielded detections of most sources. 
Through the 1\farcs5 slits, telluric emission-lines had a FWHM of 
143\kms, and spectra of unresolved sources had spectral resolution
of 76 to 95\kms.

\subsection{Continuum Imaging}

Deep, broadband imaging complements our emission-line survey because
it is more sensitive to continuum emission. Photometry can
confirm redshifts via identification of the Lyman break,  measure
star formation rates via the strength of the rest-frame UV continuum,
and constrain age via the Balmer break.  Scattering of \lya\  photons
by intergalactic hydrogen at $4.0 < z < 5.7$ attenuates the continuum
between the Lyman break and the \lya\ line, allowing only $\sim 20\% $ 
transmission (Fan \et 2002). At redshift 5.7, the \lya line 
falls in the $i^{'}$ (and  $i$ or $I$)  band, so imaging in $z$ (or IR) filters 
provides UV (optical) continuum measurements. Much of the $r^{'}$ (and
$r$ and $R$) bandpass lies below the Lyman limit.  Only filters passing 
light shortward of the Lyman limit provide good veto imaging for high-redshift
candidates.


\subsubsection{10H Field}

Taniguchi \et 2007 present  Subaru/Suprime-Cam imaging 
of the COSMOS field in the UBgVrIz broadbands and one 
narrowband NB816. The imaging redward of our emission lines
reaches $z^{'} = 25.9$ (Taniguchi \et 2007; 2\arcsec diameter 
apertures; $3\sigma$), which is deeper than the $z \approx 25.0$ 
(S. Gwyn website)\footnote{
http://www3.cadc-ccda.hia-iha.nrc-cnrc.gc.ca/community/CFHTLS-SG/docs/cfhtls.html}
imaging from the Canada France Hawaii Telescope Legacy 
Survey (CFHTLS-Field D2). Most of the emission-line objects that we find 
are not bright enough, $\ge\ 5 \times 10^{-17}$\flux, to be detected in the 
COSMOS $z^{'}$ band.

Assuming flat or redder SEDs (spectral energy 
distributions) in $f_{\nu}$, objects not detected in the Subaru $z^{'}$ 
imaging will not be detected in either the Subaru $i^{'}$ imaging or the HST 
ACS F814W observations. The latter are slightly more sensitive than the Subaru
imaging for sources less than 1\farcs0 in diameter. The bandpass of our 8190\angs\ 
filter lies just within the red limit of these filters, so these broadband filters
sample mainly attenuated continuum.  The $Bg^{'}V$ bands from COSMOS/Suprime-Cam 
are below the Lyman limit of $z=5.7$ objects, so a detection would veto 
identification of the object as a high-redshift galaxy.

%
%

\subsubsection{15H Field} \label{sec:15Himages}

Far less imaging is available for the 15H Field.  The $z^{'}$ imaging 
from LCIRS is significantly deeper than the infrared imaging. It
reaches $z^{'} = 23.5$, equivalent to $z^{'}_{AB} = 24.0$.

We obtained a new, deep image in a 3000\angs\ wide filter, WB48-78,
in 2005 March. A total of 3.5~hr of integration was obtained 
with the IMACS short camera on Magellan. The blue half of this bandpass 
lies longward of the Lyman limit, but the red half includes the \lya\ and 
Ly$\beta$ attenuated continuum. Null detections in the WB48-78 image
can therefore confirm a continuum break, but a detection does not necessarily
rule out a continuum break.

\section{RESULTS}

\subsection{Detection of Emission Lines}

Over cosmological distances, the HI opacity of the IGM 
blankets the continuum from the Lyman limit to the \lya line, 
effectively shifting the continuum break up to the \lya line (Madau 1995).
Any continuum blueward of a redshift 5.7 \lya\ line must be weak 
due to \lya\ scattering along the line-of-sight. None of the 
line-plus-continuum spectra present continuum breaks across the emission
line, so we excluded all the line-plus-continuum sources from the preliminary
list of 136 \lya\ candidates. Few of the line-only objects in the 10H field 
were detected in the Subaru narrowband image confirming our argument that 
spectroscopic surveys offer unique emission-line sensitivity.
Table~\ref{tab:class} shows the object yield for 
each of the four search runs, distinguishing line emitters with and without 
continuum in their discovery spectra.




%

\subsection{Confirmed LAEs} 

We have confirmed three LAEs among the emission-line galaxies: 
\m80+3 in the COSMOS field and \m29.5+5 and \m71-5 in the 15H field.
Confirmation requires: (1) the broad bandpass spectrum presents 
no spectral lines between $\sim 5500$ and 9500\AA\ other than the 
discovery line near 8200\AA\ and (2) the profile of this line, at 
higher dispersion, is skewed to red wavelengths, consistent with 
transmission through high-redshift IGM. In this section, we present 
spectra of these three objects, put limits on the number of LAE
candidates, and derive the survey completeness. Properties of the 
LAEs are further discussed in \S~\ref{sec:properties}.

\begin{figure}[h]
      {\includegraphics[scale=0.5,angle=-90,clip=true]{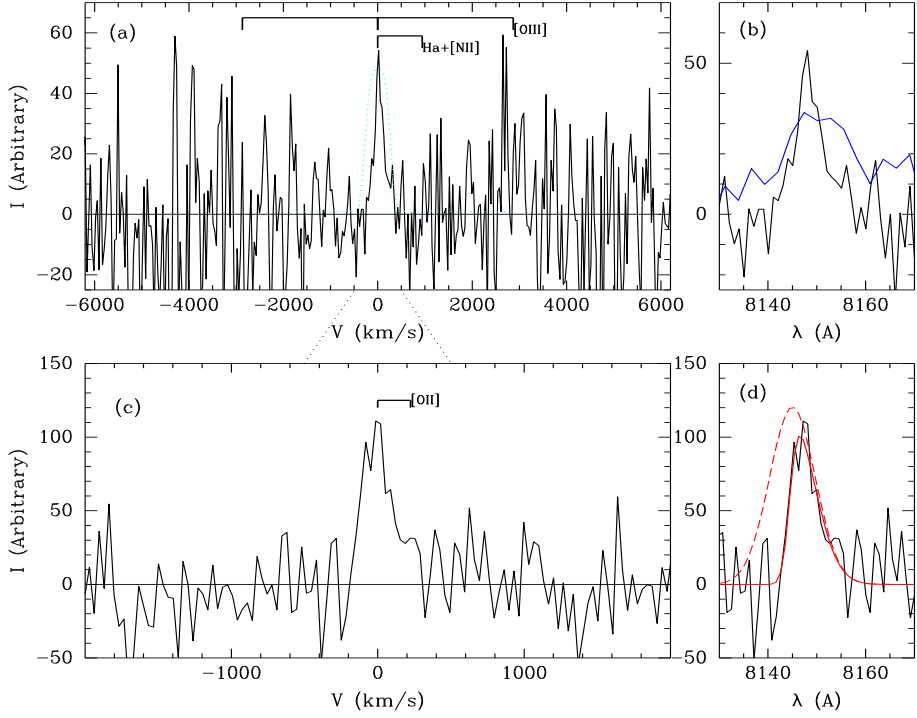} 
        }
          \caption{LAE MSDM29.5+5. 
	    The zeropoint of the velocity scale is arbitrary, defined with 
	    zero at maximum intensity for convenience. Tics mark the 
	    separations of lines that commonly appear in our spectra of 
	    foreground interlopers.
	    {\it (a):} IMACS 200-l f/2 spectrum taken in 
	    2005 June. The line is narrow compared to the profile of a 
	    sky line, shown by the light-blue, dotted line.
	    {\it (b):} Spectrum from {\it (a)} on a wavelength scale.
	    Also shown is an IMACS 150-l f/2 spectrum taken taken in 2006 
	    June (thick, blue line).
    {\it (c):} Higher resolution, Keck/LRIS spectrum.
	    The line is asymmetric and skewed to the red.
	    {\it (d):} Keck/LRIS spectrum and fit. The model
	    is a Gaussian (400\kms FWHM) with flux blueward of the 
	    peak set to zero and then convolved with another Gaussian 
	    representing instrumental broadening (110\kms).
          }
           \label{fig:msdm29.5+5} \end{figure}

Lyman-alpha-emitter MSDM29.5+5 from the 15H field shows the 
classic red-skewed line asymmetry of a LAE in the high-resolution,
Keck spectrum; see panel (c) of Figure~\ref{fig:msdm29.5+5}.  
The line width is 235\kms FWHM. 
The low resolution spectrum, panels (a) and (b), is not 
sufficient to recognize the asymmetry. The line 
in the 200-l spectrum is  narrower than the overlaid sky line,
indicating a source size considerably smaller than the slit.
%
The attenuated \lya\ flux is  $\sim 1.8\pm0.2 \times 10^{-17}$\flux\
in the 2005 May discovery spectrum. The systematic errors in the
flux calibration may be more uncertain than this statistical error
due to the observing conditions described in  \S~\ref{sec:observations}. 
The observed-frame equivalent width is at least 11\angs\ based
on the $3\sigma$ upper limit on the continuum in the search spectrum.
This object is outside the field of the LCIRS image.

\begin{figure}[h]
      {\includegraphics[scale=0.5,angle=-90,clip=true]{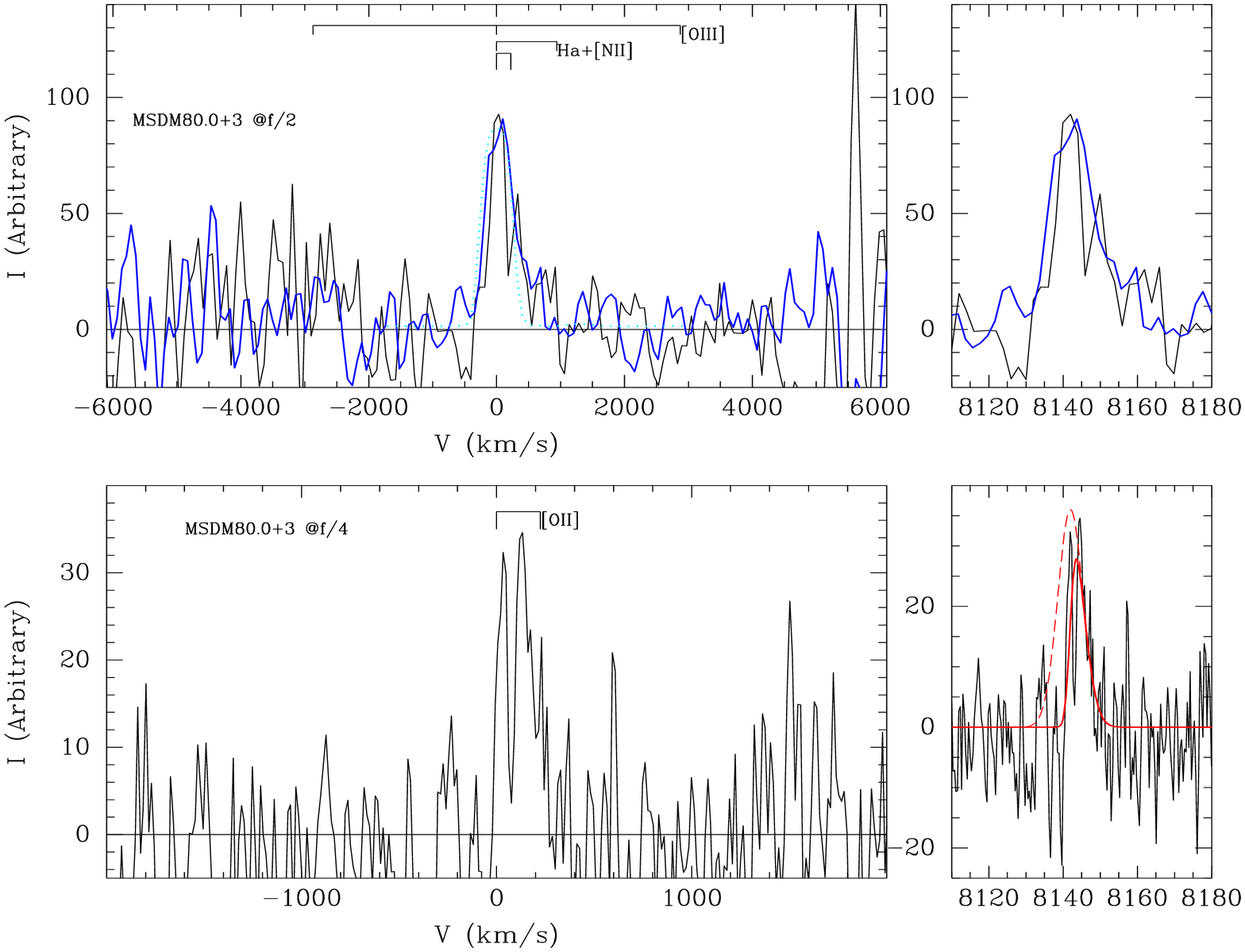} 
        }
          \caption{LAE MSDM80+3. Velocity scale, tic marks, and arc line as 
	    in \fig~\ref{fig:msdm29.5+5}. {\it a,b:} IMACS f/2 spectra from
	    2005 May (thin, white line) and 2007 February (thick, blue line).
	    {\it c:} IMACS f/4 spectrum. 
	    {\it (d):} IMACS f/4 spectrum and fit. 
	    The model
	    is a Gaussian (330\kms FWHM) with flux blueward of the 
	    peak set to zero and a 150\kms wide Gaussian 330\kms
	    redward of the first. These components were
	    convolved with another Gaussian 
	    representing instrumental broadening (150\kms).
          }
           \label{fig:msdm80+3} \end{figure}

Figure~\ref{fig:msdm80+3} shows LAE MSDM80+3, discovered in the 2004 search 
of the 10H field. The broad bandpass spectrum shows a slight red asymmetry 
of questionable significance.  The higher resolution spectrum shows multiple
line components spread over about $400 $\kms FWHM.
The \lya flux is $1.4\pm0.2 \times 10^{-17}$\flux.  The $3\sigma$ 
limiting $z^{'}$ flux (for 2\arcsec\ diameter apertures) gives a lower limit 
on the observed-frame equivalent width of 144\angs.

We believe \m80+3 is the same object as \#55 detected by Muramyama \et 
(2007) using Subaru/Suprime-Cam
narrowband imaging. The declinations agree to 0\farcs2. Their right 
ascension is nearly 1\farcs0 west of ours, consistent with the 
zeropoint shift we found relative to the Subaru image in 
\S~\ref{sec:observations}. They estimate a significantly higher line 
flux of $2.8 \pm 0.4 \times 10^{-17}$\flux.  The discrepancy could result 
from attenuation of the source by our slit.  Line fluxes from
narrowband imaging are, however, subject to uncertainties about redshift,
filter transmission, and continuum level. Since no continuum is detected
from MSDM80+3 in broadband $i^{'}$ or $z^{'}$ images (Taniguchi \et 2007) 
and we find a wavelength near the center of the filter bandpass, we see
no obvious reason for the imaging result to be in error. We quantify
our slit losses on a statistical basis in \S~\ref{sec:complete} and
include them in modeling the number counts in \S~\ref{sec:counts}.



\begin{figure}[h]
      {\includegraphics[scale=0.5,angle=-90,clip=true]{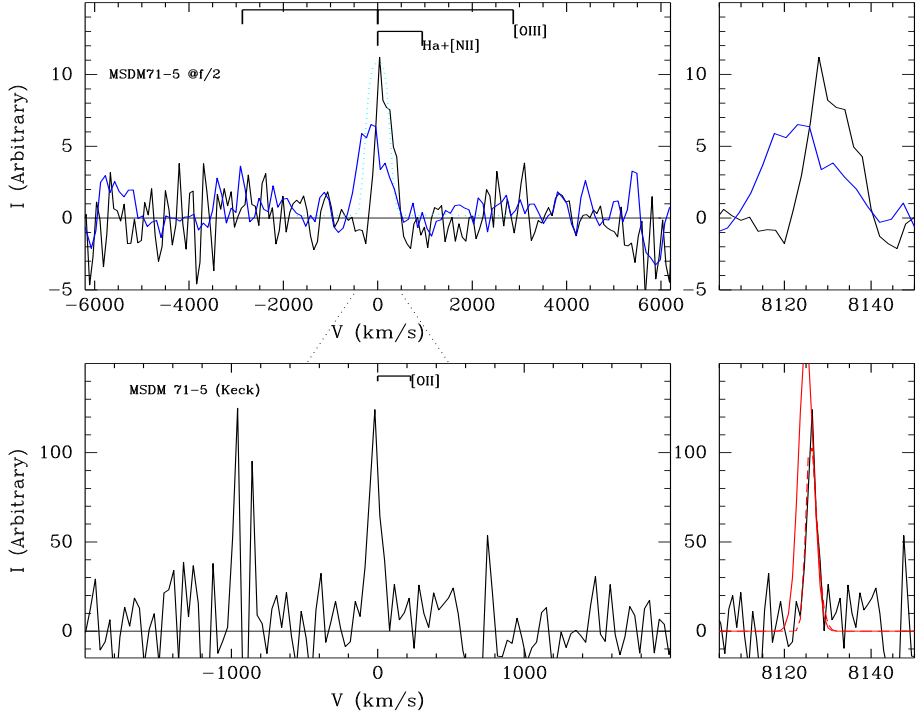}
         }
      \caption{LAE MSDM71-5. Velocity scale, tic marks, and arc line as 
	    in \fig~\ref{fig:msdm29.5+5}. {\it Top:} IMACS f/2 spectra with
	    the 200-l grism in 2004 July (thin, white line) and the 150-l
	    grism in 2006 June (thick, blue line). {\it Bottom:} IMACS f/4
	    spectrum. 	    The modelis a Gaussian (150\kms FWHM) 
	    with flux blueward of the  peak set to zero and then
	    smoothed  by an instrumental response of width 75\kms. 
}
           \label{fig:msdm71-5} \end{figure}

Another line-emitter, \m71-5, marginally meets our criteria to 
be labeled ``confirmed.'' The shape of the line-profile leaves
some ambiguity.  Figure~\ref{fig:msdm71-5} shows follow-up spectra of 
\m71-5 taken with the 200-l and the 150-l grisms. Even at such low 
resolution, both of these spectra hint at a broad, red wing.
Curiously, our high resolution Keck, LRIS spectrum shows a narrow line. 
It is highly unlikely that different objects were observed with the
two telescopes because the line appears at the same wavelength in all 
3 spectra. The B and z LCIRS images show no objects near the position of 
\m71-5. We detect an object slightly less than 1\farcs5 to the
southeast in our deeper WB48-78 image; but, as explained in 
\S~\ref{sec:15Himages}, a weak detection in this band does 
not veto a line identification as \lya.

The coordinates, approximate wavelength, and line flux of
these confirmed LAEs are given in Table~\ref{tab:properties}.
We also tabulate the properties of emission line objects 
for which identification of the line as \lya\ has not
been ruled out, as we explain in the next section.



\subsection{Additional LAEs?}



We now ask whether our selection criteria may have missed some LAEs.
Our follow-up observations were unusually complete. Low resolution, 
follow-up spectroscopy has been attempted for all but three {\it
line-only} objects in each field. Should we have rejected all the 
{\it line-plus-continuum} objects?  Many line-plus-continuum objects 
were in fact used to fill our follow-up masks. All their spectra obtained 
to date have been identified as foreground objects (see Table~\ref{tab:class}), 
so our initial continuum cut appears to be valid.  The only sources that
allow further speculation are the un-recovered targets and the galaxies
that present a single emission line, i.e. just the discovery line near 8200\AA,
in their broad bandpass follow-up spectrum.

Table~\ref{tab:class} reveals that in the 10H and 15H fields, 13 and 7 
candidates, respectively, were not recovered by follow-up observations.  
Some candidates with line fluxes near the detection limit,  derived in
\S~\ref{sec:complete}, will inevitably turn out to be spurious.
Review of the stacked, search data, led us to believe the majority of 
such sources were genuine, however. We attribute some failed recoveries, 
particularly for the 2004 sources, to the relative astrometric errors 
described in \S~\ref{sec:f2spectra}. The fraction of LAEs among these
objects is presumably as small as it is for the full sample, so we
estimate 0 - 1 LAEs were missed.  Our follow-up observations are
also biased against detecting the faintest LAEs in our sample. Time and 
weather limitations resulted in broad bandpass spectroscopy with
half the exposure time of the initial search. These objects may include 
a higher, but unknown, fraction of LAEs.

Another set of 14 candidates (7 in the 10H  field and 7
in the 15H field) are listed in Table~\ref{tab:properties}.
These objects present a single emission-line, at $\sim 8200$\angs,
in their broad bandpass spectrum. Higher resolution spectra, had
they been obtained for these objects,  would differentiate between
LAEs, foreground [OII] emitters at $z=1.2$, and extremely faint
[OIII] 5007 emitters -- i.e. those with undetectable [OIII] 4959 and 
\Hb\ lines. Among the objects observed at higher-dispersion, the yield of LAEs 
to [OII] emitters varied widely among fields, making it difficult to
predict the LAE fraction among the single-line sources.  We gained
additional insight from photometric redshifts from COSMOS (Mobasher et al. 2007)
in our 10H field.  The continuum emission from our candidates is typically 
fainter than the $i^{'} < 25$ completeness limit.  We did investigate
the redshifts of foreground galaxies near the sightline to each line
emitter, however, with the idea that associated satellite galaxies or 
HII regions might be the source of the detected line emission.
We find foreground galaxies surprisingly close to many of the 
single-emission-line galaxies, and their photometric redshifts often
allow  [OII] emission to be shifted into our search bandpass. 
For readers interested in the exact accounting, sections \ref{sec:10h} 
and \ref{sec:15h}  review the single-emission-line objects 
individually in the 10H and 15H fields, respectively.


\subsubsection{Single-Emission-Line Candidates in the 10H Field} \label{sec:10h}

\begin{figure}[h]
      {\hbox { \includegraphics[scale=0.19,angle=-90,clip=true]{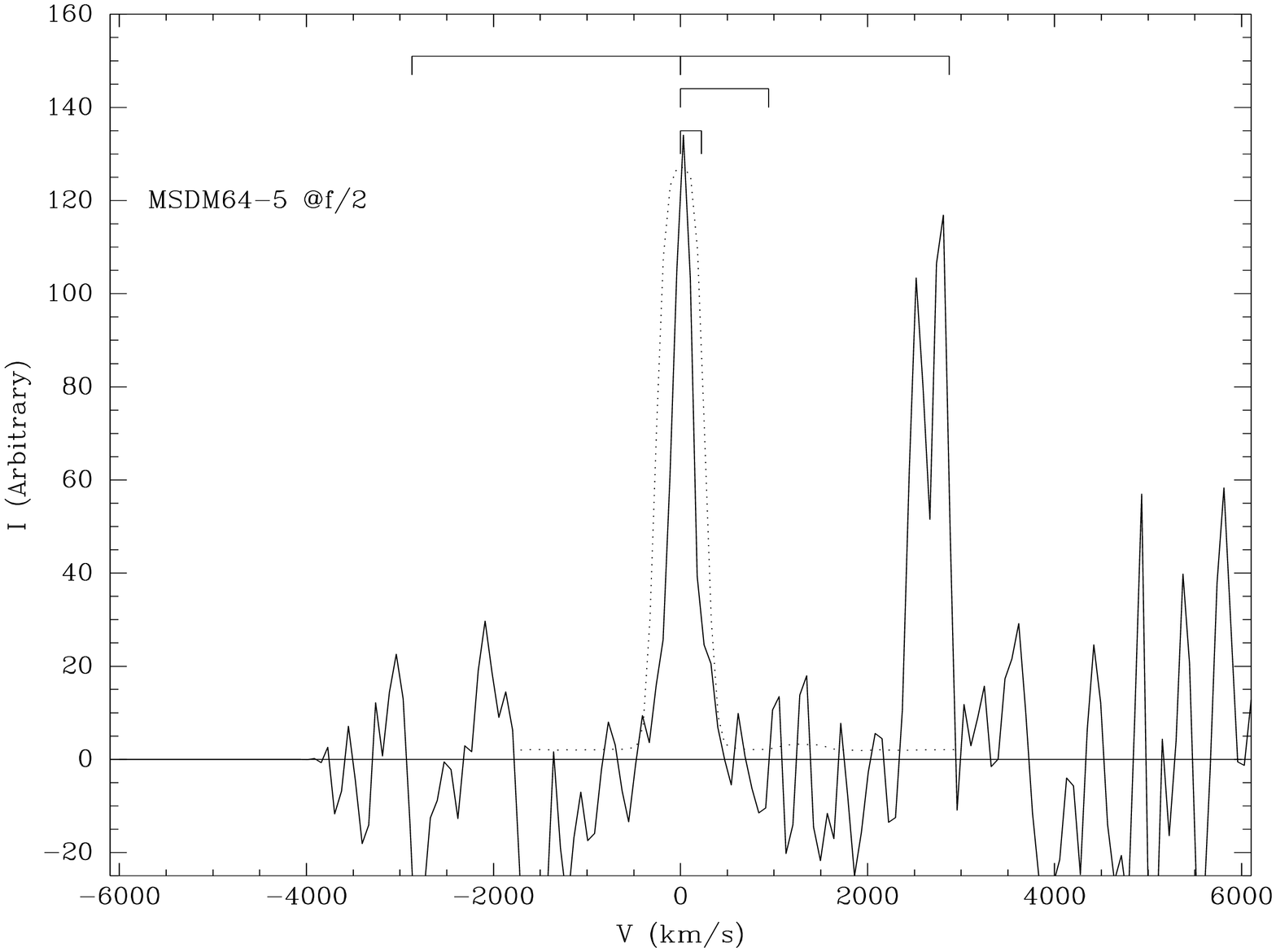} 
	\includegraphics[scale=0.19,angle=-90,clip=true]{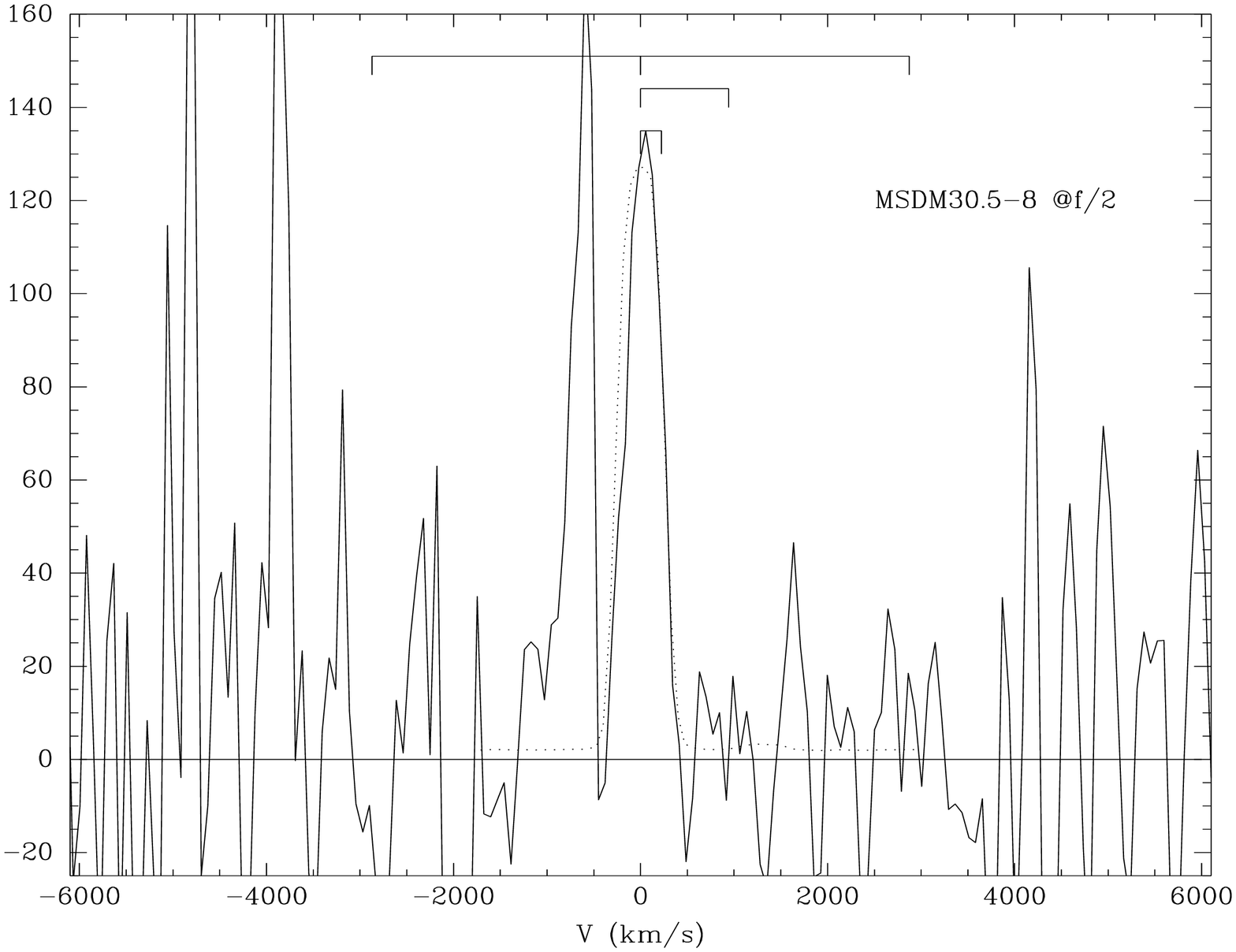} 
          }}
      {\hbox { \includegraphics[scale=0.19,angle=-90,clip=true]{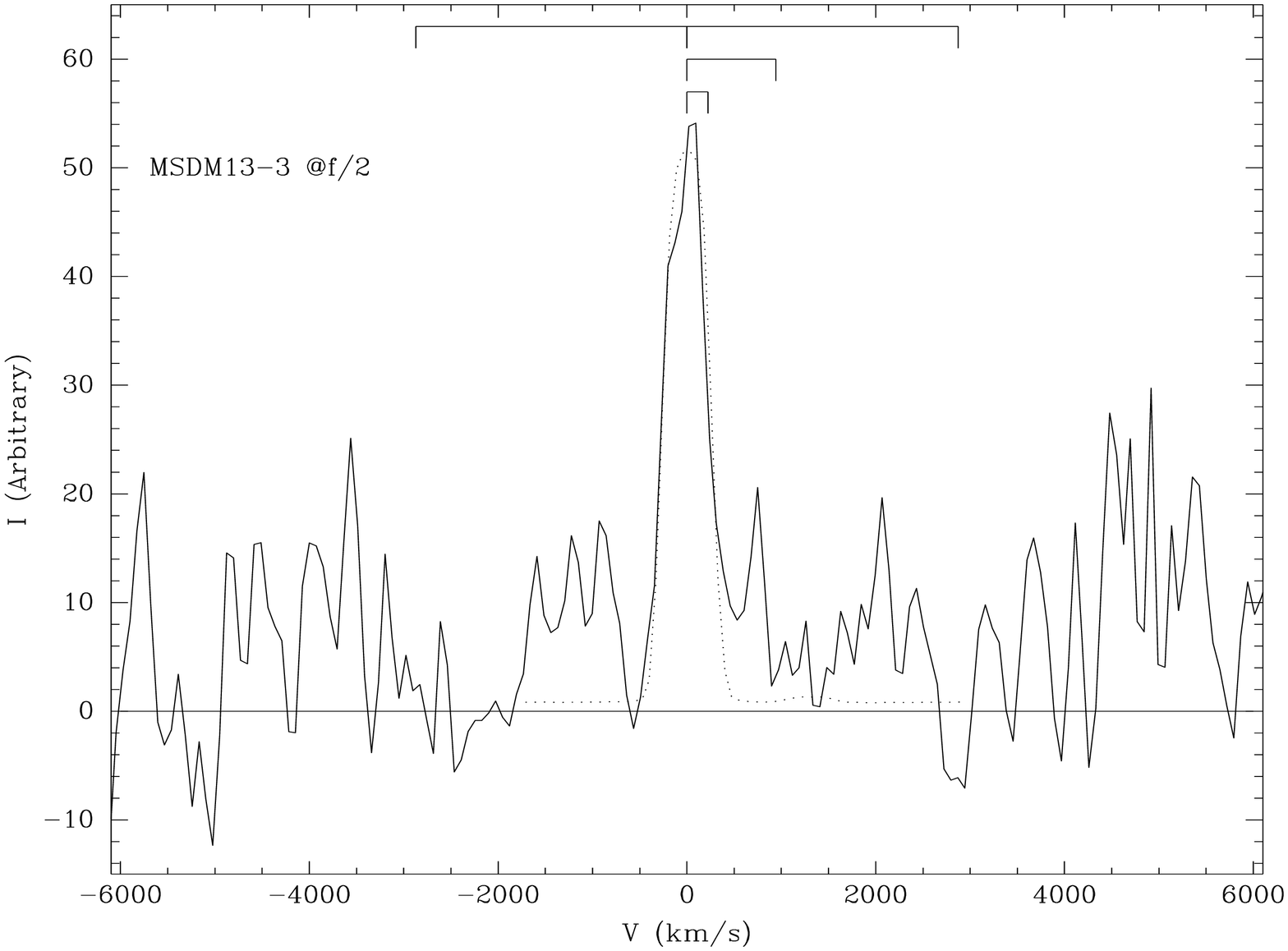}
	\includegraphics[scale=0.19,angle=-90,clip=true]{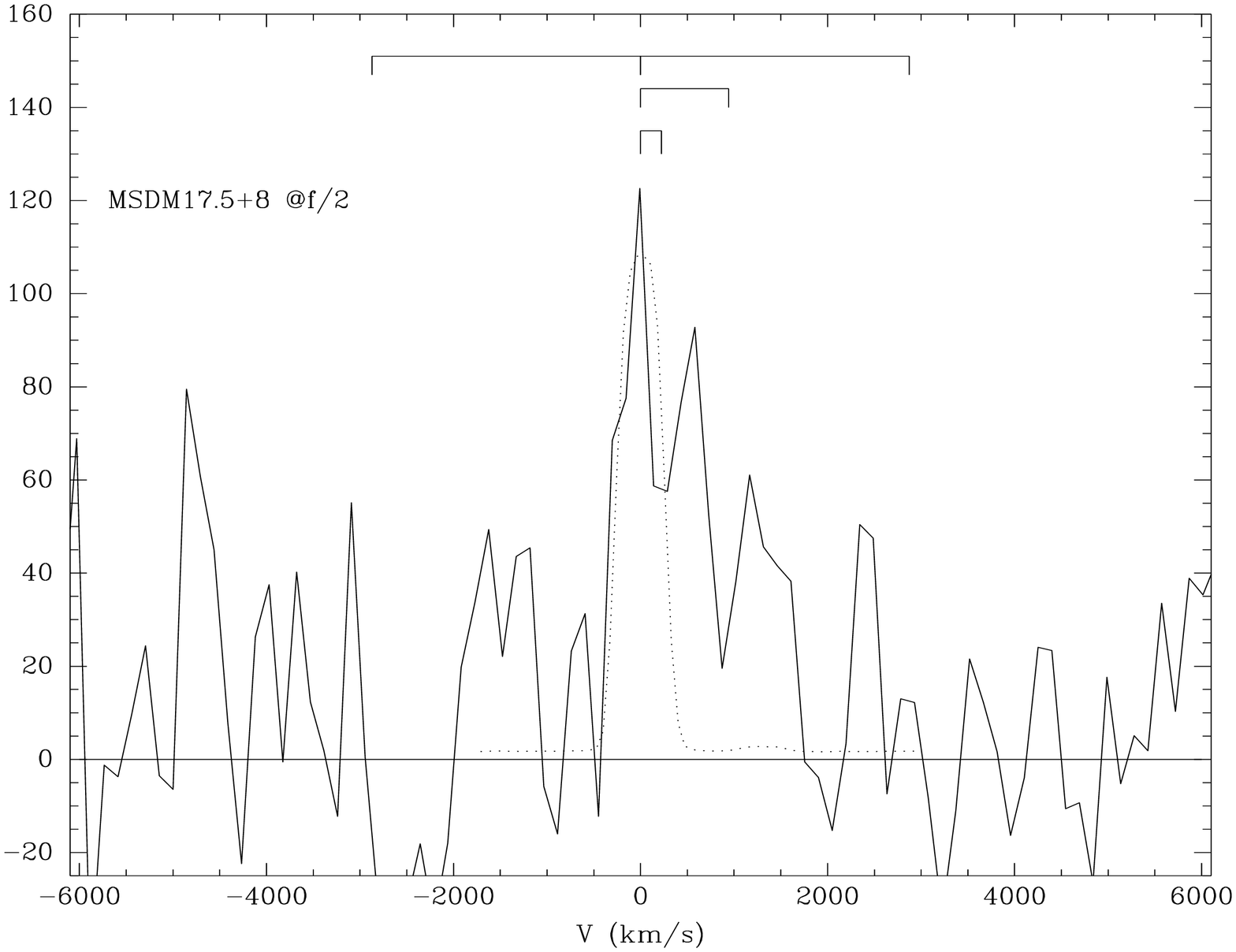}
          }}
      {\hbox { \includegraphics[scale=0.19,angle=-90,clip=true]{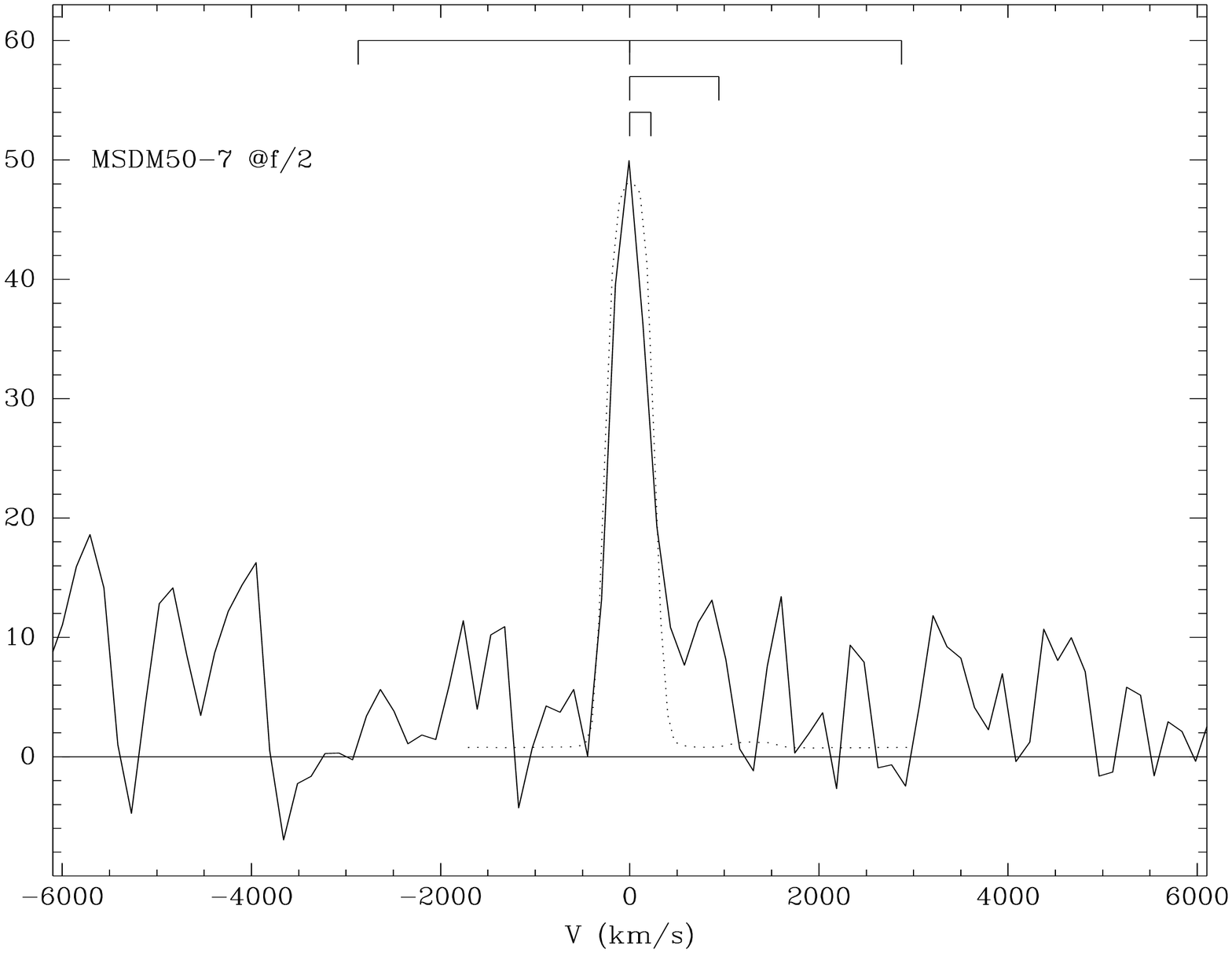}
	\includegraphics[scale=0.19,angle=-90,clip=true]{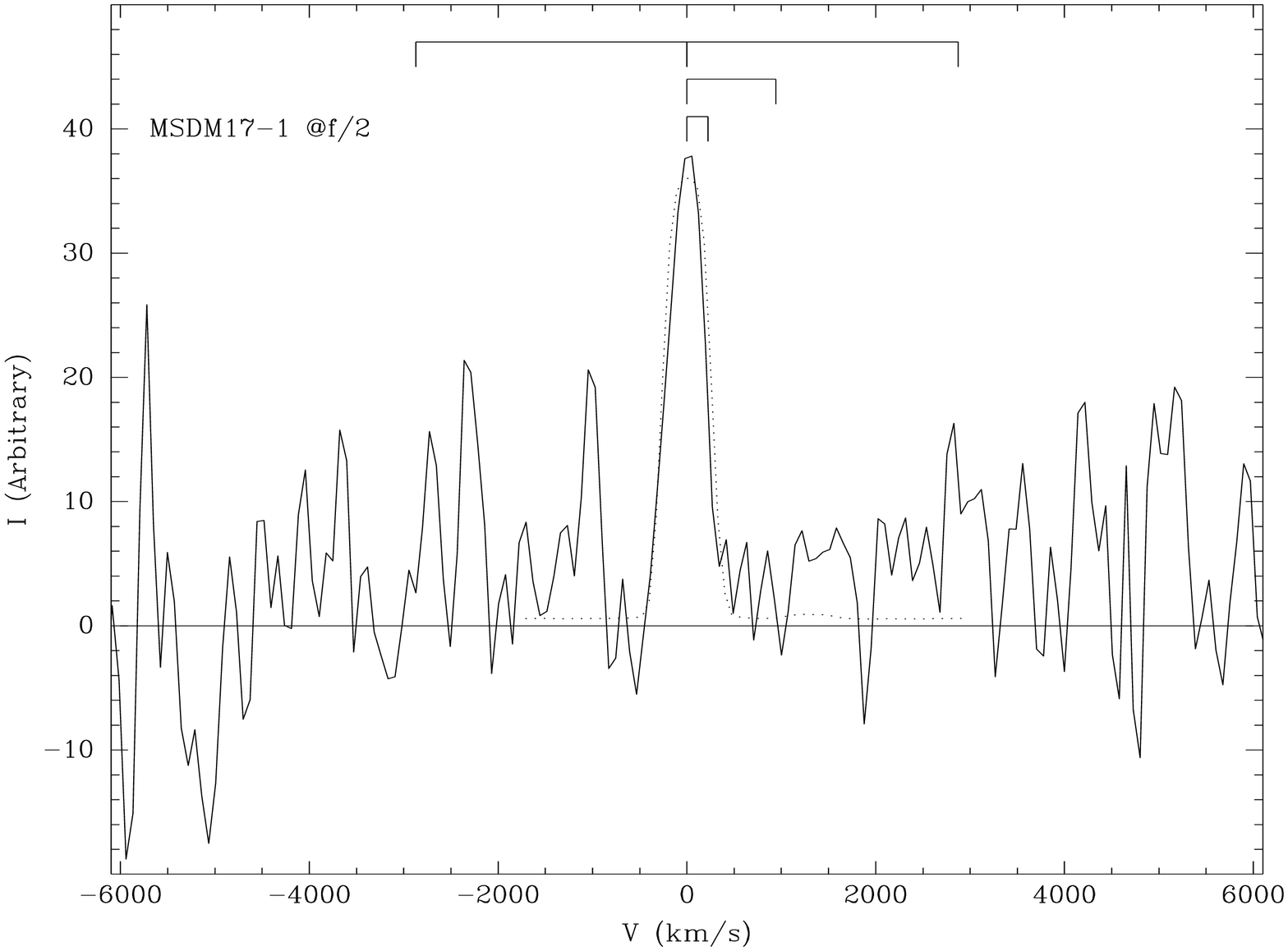} 
         }}
      {\hbox {\includegraphics[scale=0.19,angle=-90,clip=true]{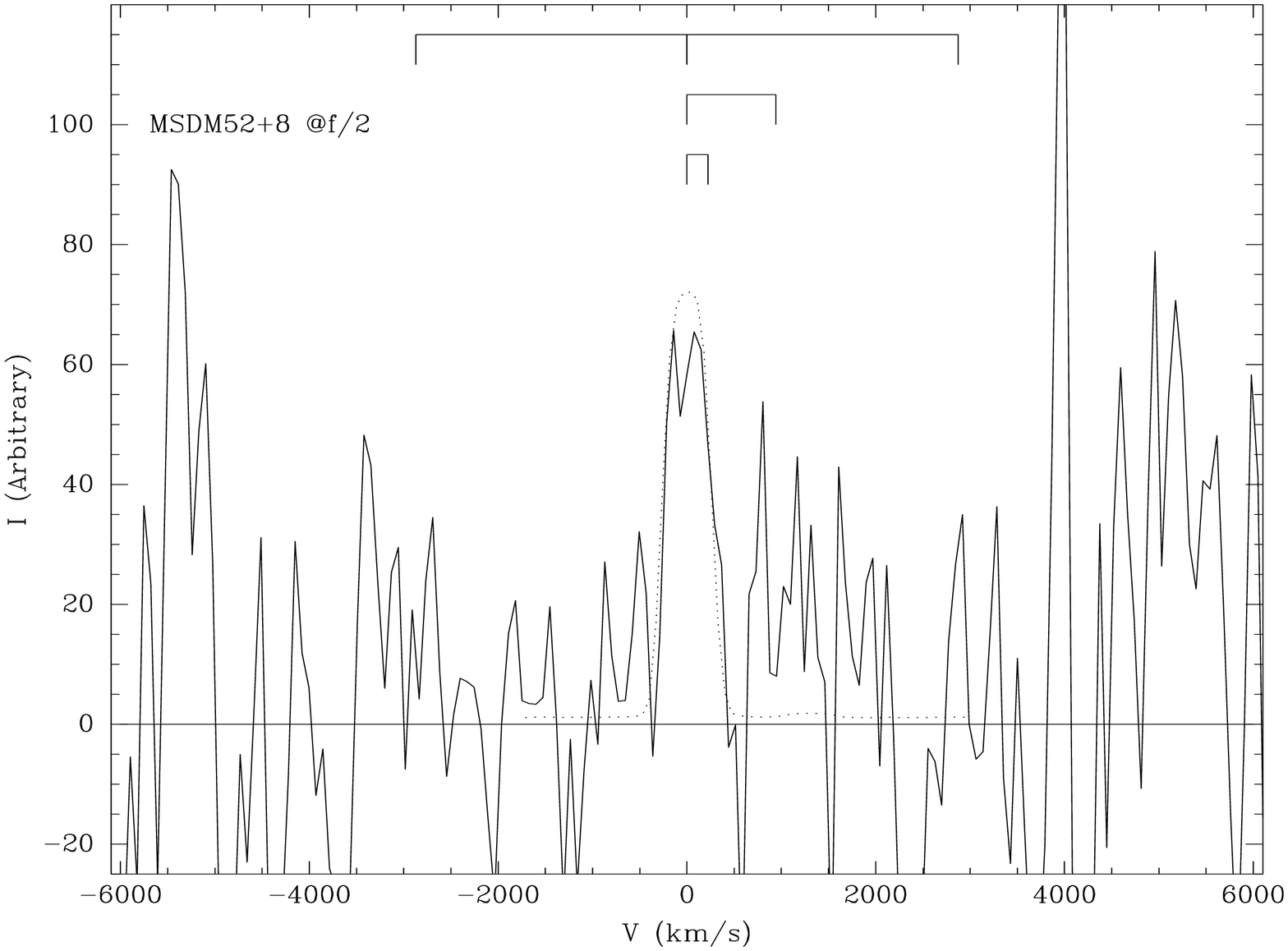}
         }}
          \caption{Low-resolution (f/2) spectra of 10H Field LAE 
	    candidates. The objects shown have a single line in the
	    observed-frame 5000-9000\AA\ bandpass. This inset shows
	    no lines are seen at the spacing (tic marks) of the 
	    foreground contaminants [OIII]5007,4959 or \Ha + [NII]6584.
	    Comparison to the  profile (dotted line) of a source that 
	    fills the slit shows the lines are unresolved, and higher
	    resolution spectroscopy is required to show line asymmetry
	    and/or identify the [OII] 3726,29 doublet (small tics).
          }
           \label{fig:10h_single} \end{figure}

The low-resolution line profiles of the seven single-line candidates
in the 10H Field are shown in Figure~\ref{fig:10h_single}.  The line
profiles of \m52+8,  \m17.5+8, and \m13-3 present significant skew. 
The spacing of the two peaks in the \m52+8 spectrum is consistent with [OII] 
emission; however, the bluer of the two emission peaks is stronger, which 
would imply a  high electron density if the line is indeed [OII]. Electron
densities in typical HII regions are lower than this, so this line can
only be [OII] if it is an AGN (Active Galactic Nucleus), 
which will rarely be the case.  In contrast,
the profile shape of \m13-3 is consistent with a blended [OII] doublet in
the low-density limit.  These spectra lack the resolution required to
reveal the asymmetry of many LAEs, such as \m29.5+5 in \fig~\ref{fig:msdm29.5+5},
so a larger fraction of the single-line sources may actually have 
asymmetric line profiles.

To identify more interlopers, we searched the photometric redshift catalog of
Mobasher et al. (2007) near the positions of our emission-line objects. The 
density of objects in the Subaru (UBgVrIz) imaging is such that a foreground 
galaxy brighter than the i=25 detection limit lies within 4\farcs5 (or 
9\farcs0) of a randomly chosen position 51\% (or 92\%) of the time.  The 
projected distance to the nearest redshift 1.2, [OII]-emitting galaxy is 
larger. In the Mobasher \et catalog, we find a galaxy consistent with
being at redshift 1.2  within 4\farcs5 (or 9\farcs0) of a random sky position  
only 3.5\% (or 11\%) of the time.  In light of these statistics, if the single-line
objects are primarily LAEs, then very few (if any) will be projected 
near a galaxy with a photometric redshift near 1.2.

As expected, no foreground continuum sources are detected near our one 
confirmed LAE in the 10H field, \m80+3. The closest galaxy whose photometric 
redshift is consistent with [OII] emission in our filter lies 9\farcs9 to the 
northeast. Our position for \m64-5 is 2\farcs1 northwest of the center of a 
foreground spiral galaxy with $z_{phot} = 0.90$.  The 95\% confidence interval, 
$0.82 < z < 1.02$, allows \Hg (and plausibly a faint [OIII] 4363 line) to fall
in our search band. The problem with this association is that \Hb and [OII] 
were not detected in the broad bandpass spectrum.

The other six single-line galaxies in the 10H field are likely foreground
galaxies. All of them lie less than 3\farcs5 from a galaxy plausibly at 
redshift $\sim 1.2$.  The associations are particularly close, in both 
angular separation and probable redshift, for \m30.5-8, \m17.5+8, and \m17-1.  
The separation is larger, 2\farcs1 (17.1~kpc at $z=1.2$), for \m52+8, but much 
of this discrepancy is in the direction of our astrometric (zero-point) error.  
We argued that \m13-3 is likely an [OII] emitter based on its line profile.
Both \m13-3 and \m50-7 are very close to faint, $i > 25$, galaxies,
which are known to be foreground owing to B-band detections.  Theses offsets
are small enough to be astrometric errors, and we tentatively assign 
our emission-line detection to the faint objects.  These $i > 25$ galaxies 
may be associated with the brighter, cataloged galaxies (near redshift 1.2)
a few arcseconds away. Further work will be required to understand their 
physical nature.







\subsubsection{Single-Emission-Line Candidats in the 15H Field} \label{sec:15h}

\begin{figure}[h]
      {\hbox { \includegraphics[scale=0.19,angle=-90,clip=true]{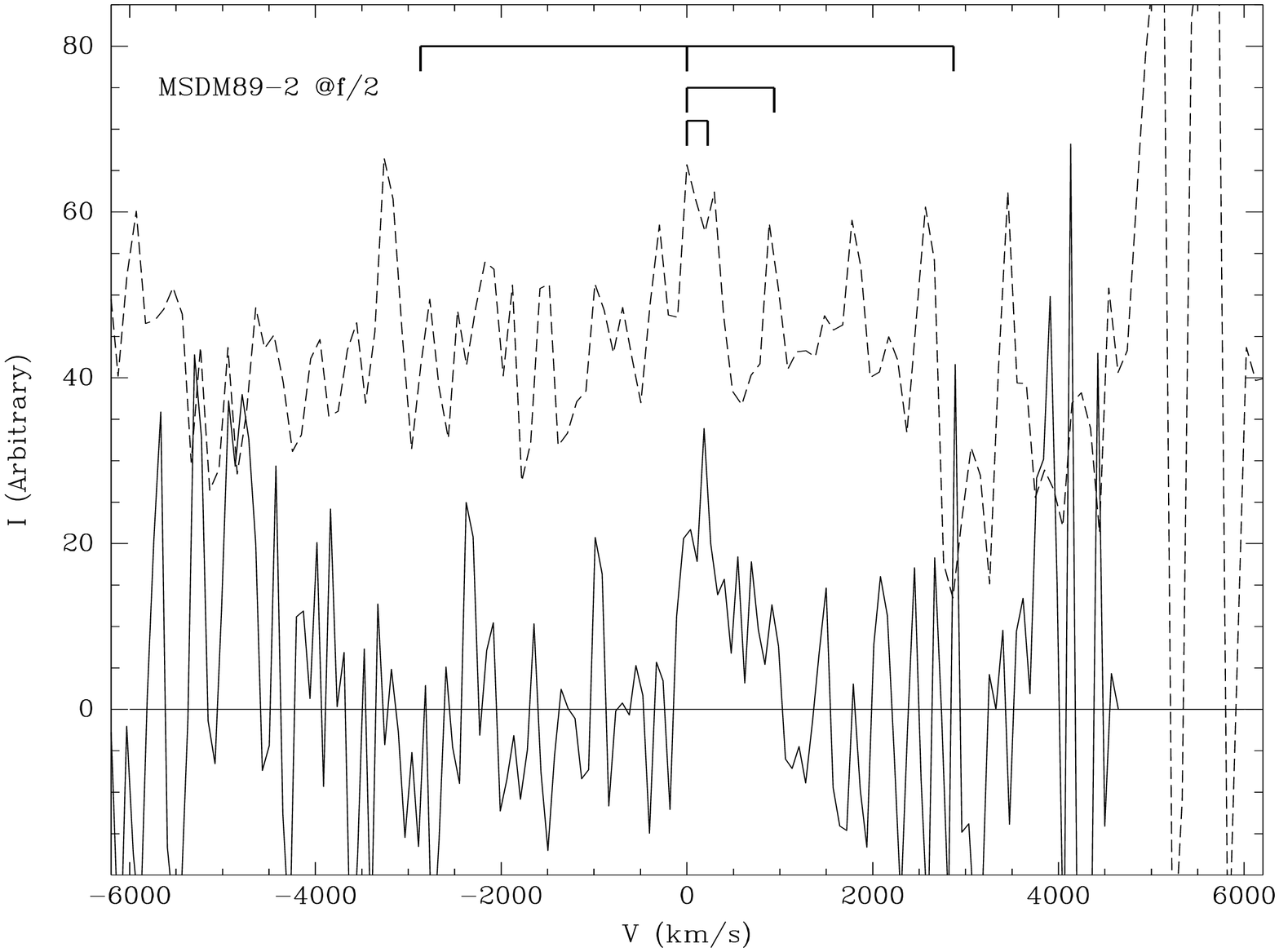} 
	\includegraphics[scale=0.19,angle=-90,clip=true]{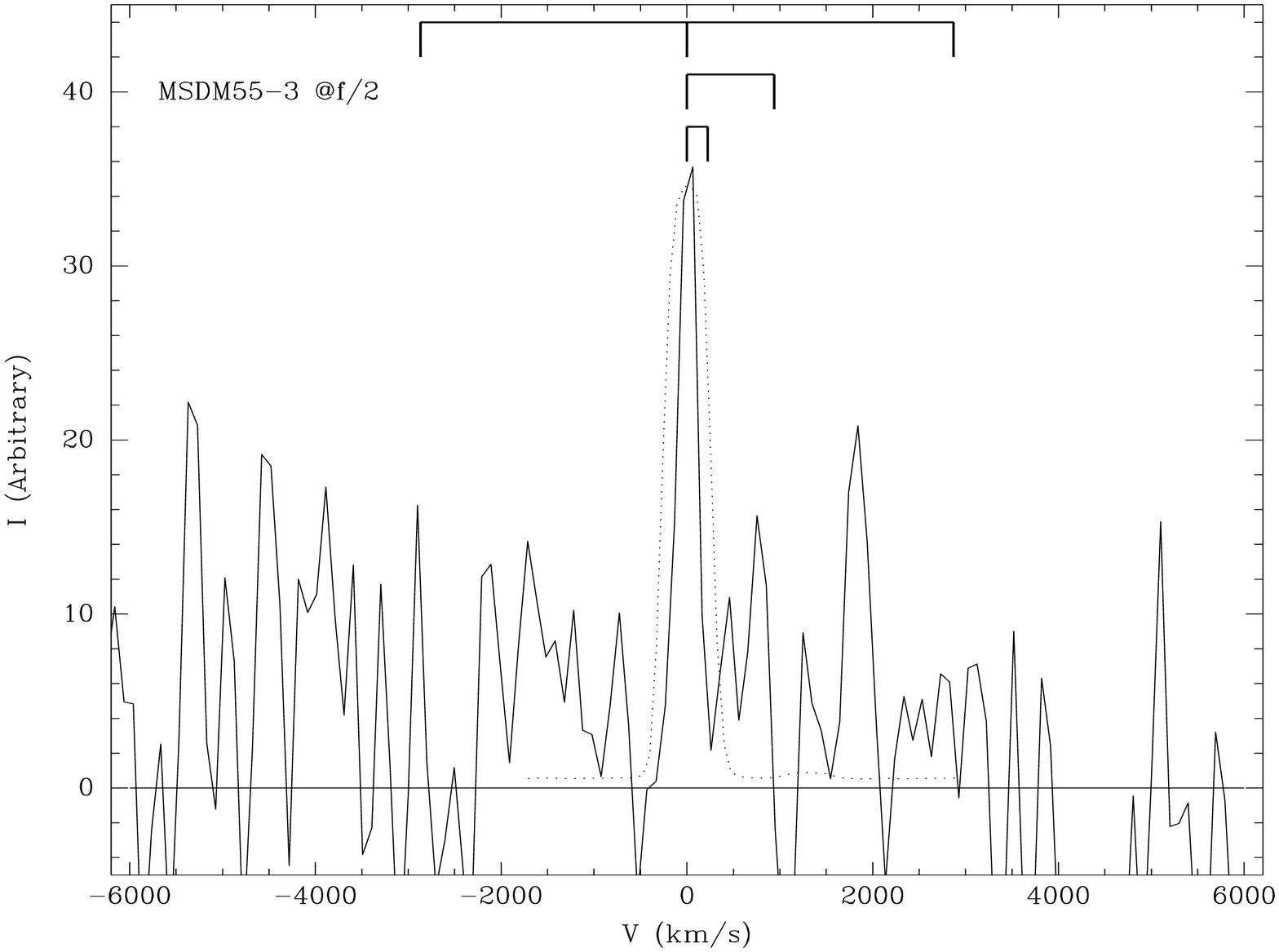} 
          }}
      {\hbox { \includegraphics[scale=0.19,angle=-90,clip=true]{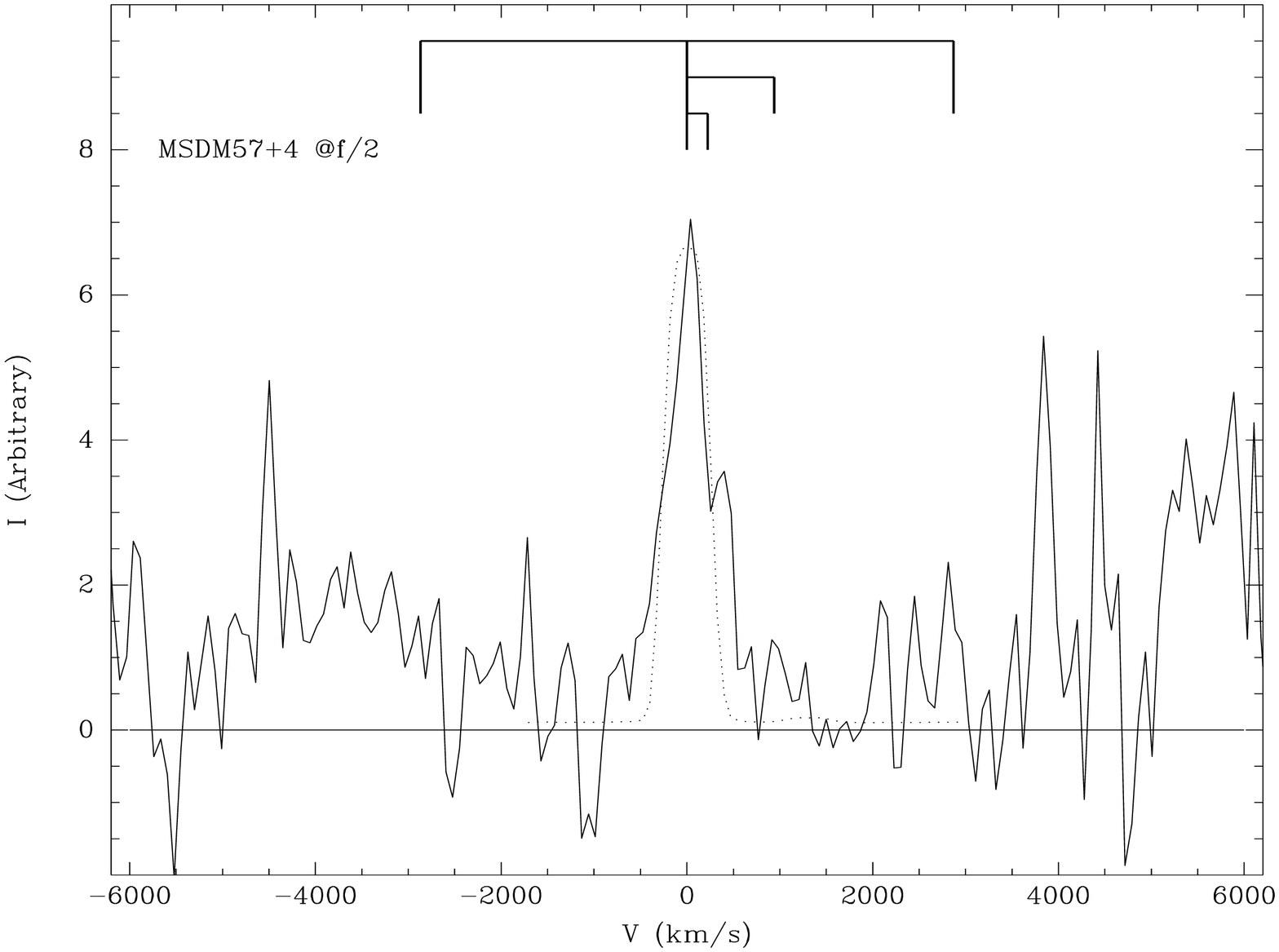}
	\includegraphics[scale=0.19,angle=-90,clip=true]{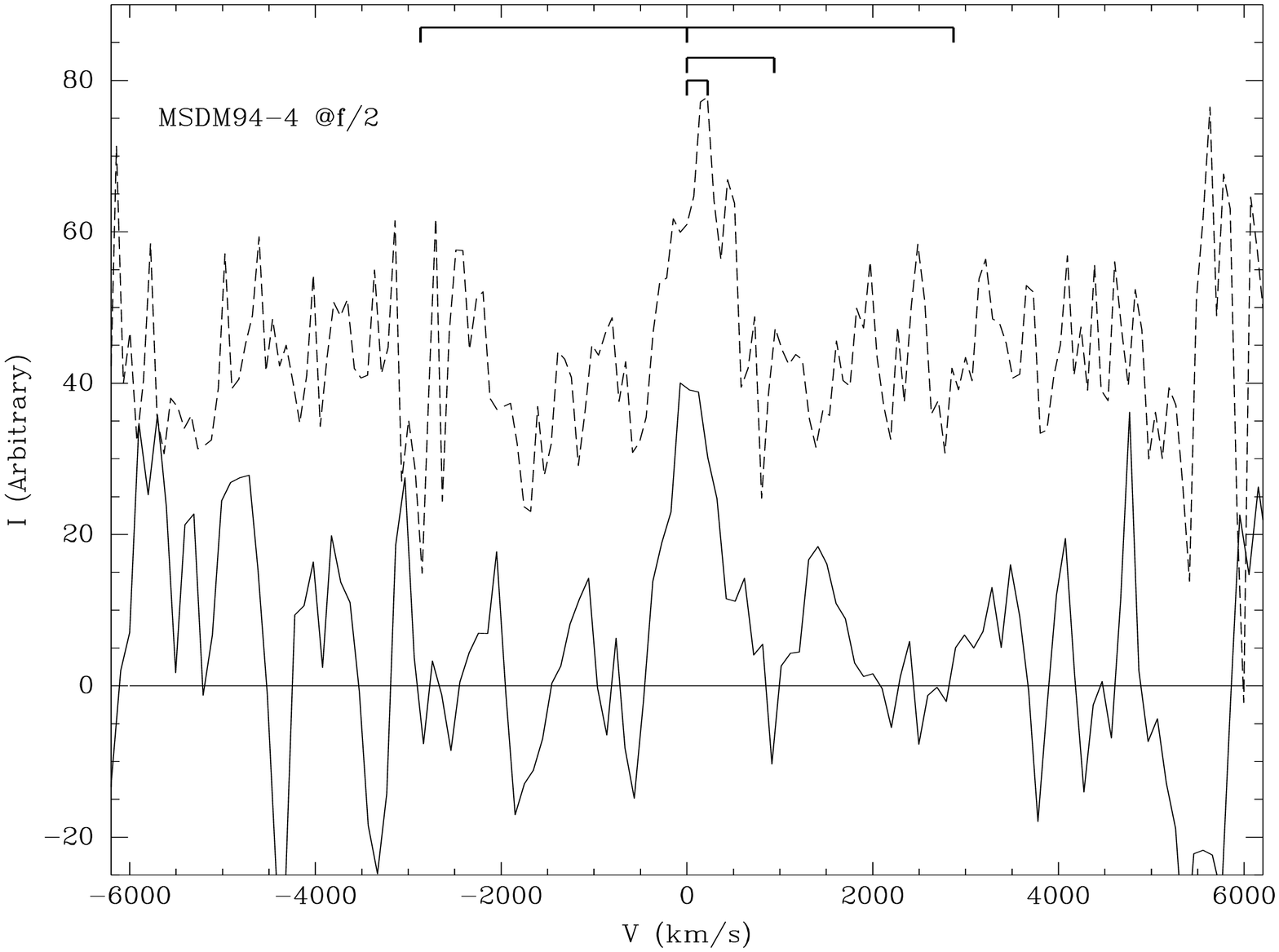}
          }}
      {\hbox { \includegraphics[scale=0.19,angle=-90,clip=true]{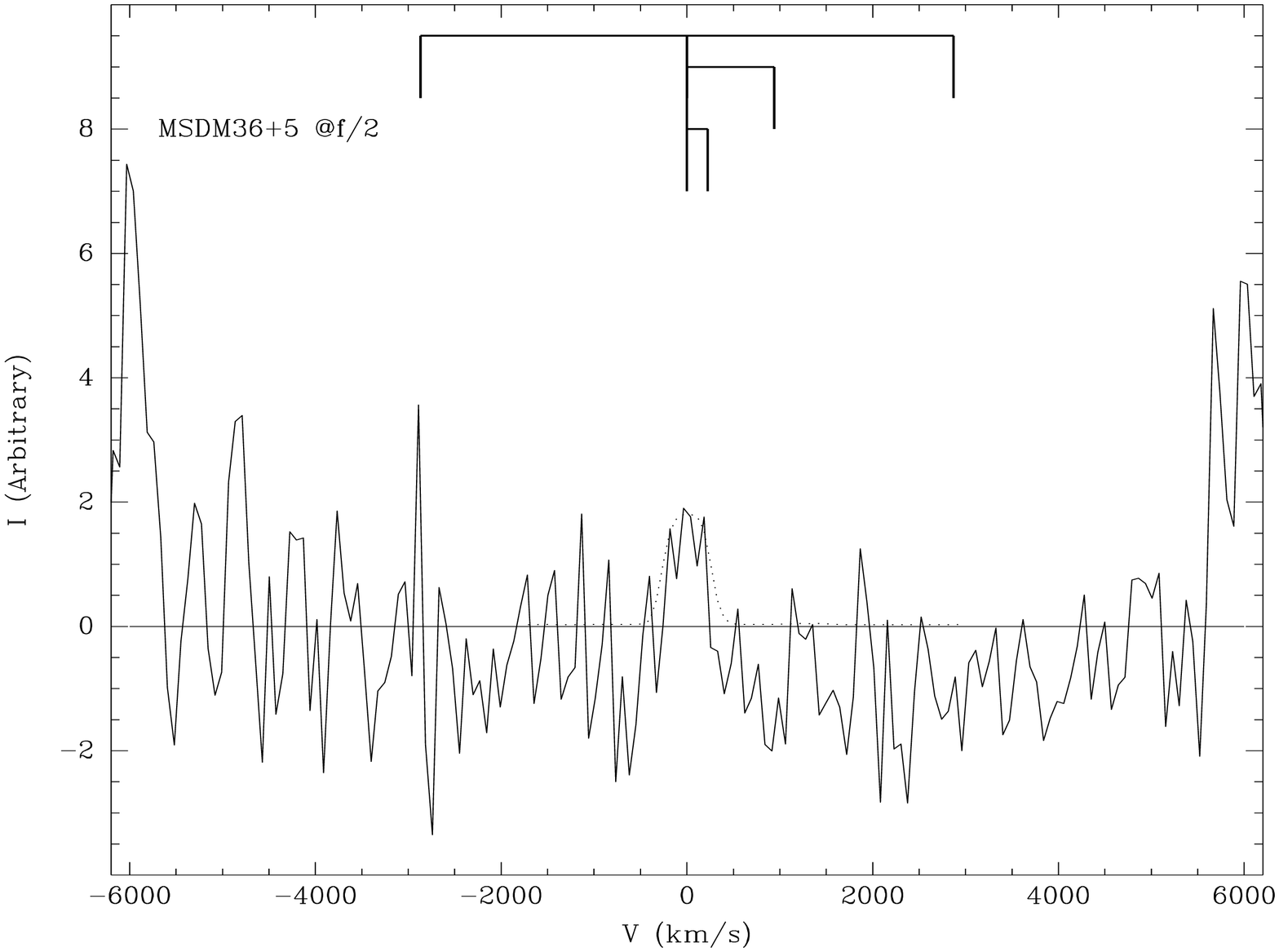}
	\includegraphics[scale=0.19,angle=-90,clip=true]{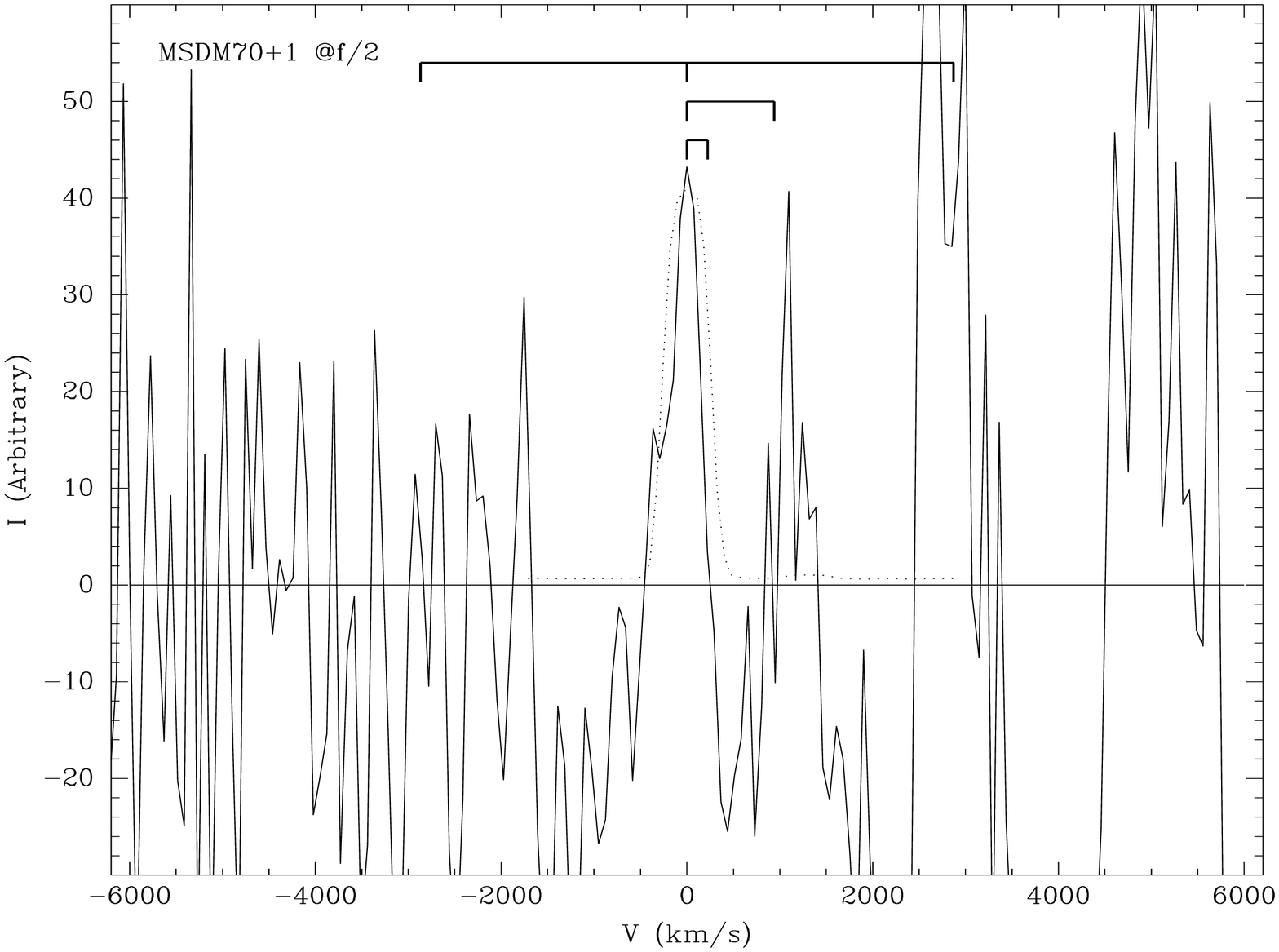} 
         }}
      {\hbox {\includegraphics[scale=0.19,angle=-90,clip=true]{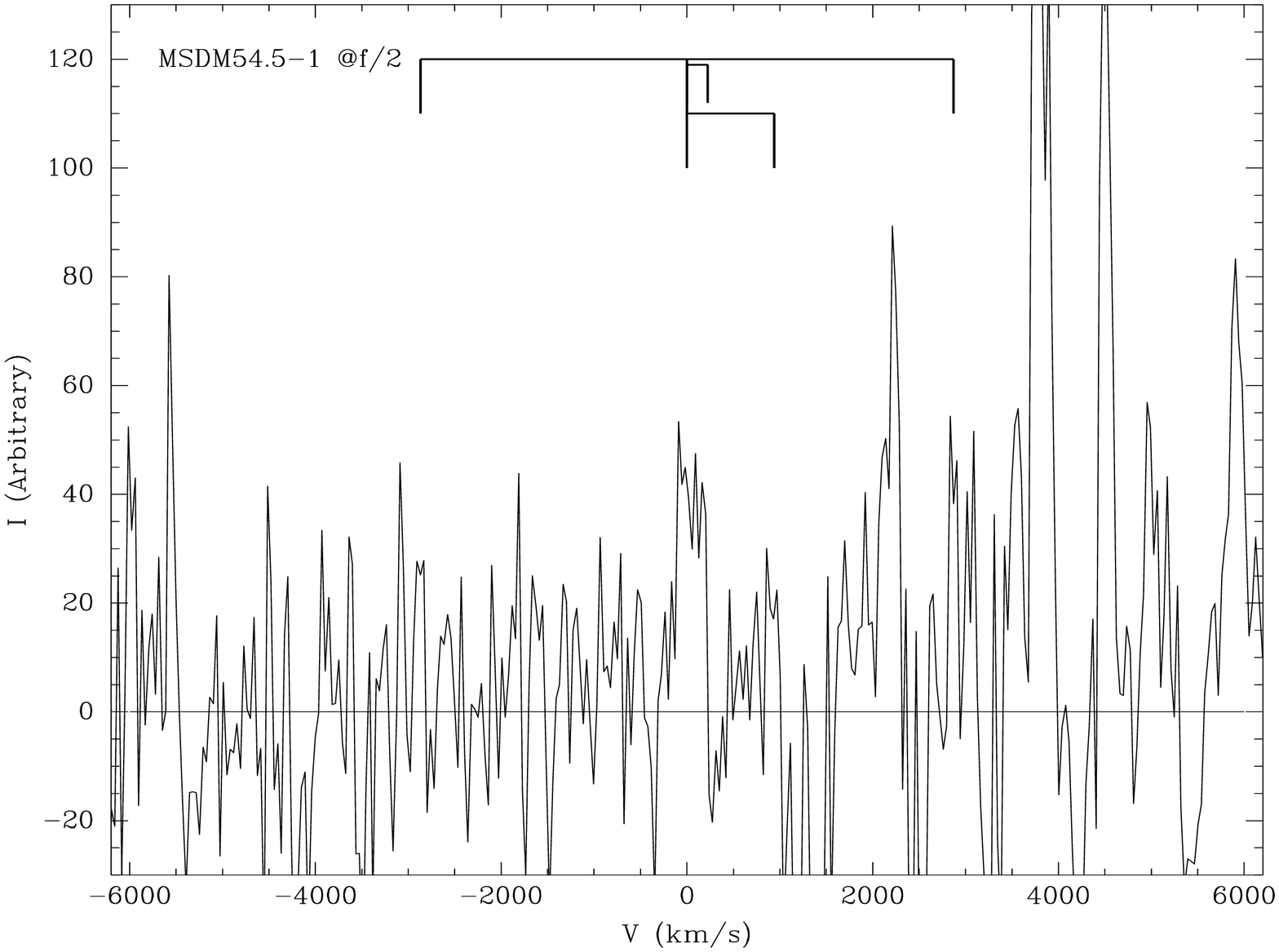}
         }}
          \caption{Low-resolution IMACS spectra of 15H Field LAE 
	    candidates. The objects shown have a single line in the
	    observed-frame 5000-9000\AA\ bandpass. This inset shows
	    no lines are seen at the spacing (tic marks) of the 
	    foreground contaminants [OIII]5007,4959 or \Ha + [NII]6584.
	    Comparison to the  profile (dotted line) of a source that 
	    fills the slit shows the lines are unresolved, and higher
	    resolution spectroscopy is required to show line asymmetry
	    and/or identify the [OII] 3726,29 doublet (small tics).
          }
           \label{fig:15h} \end{figure}

In addition to \m29.5+5 and \m71-5, we identified 7 single emission-line 
sources in the  15H field. Their f/2 spectra are shown in \fig~\ref{fig:15h}.
The line profiles of \m57+4, and \m70+1 
show some sign of asymmetry at low resolution. We interpret the blue
skew of the \m70+1 profile as evidence of the [OII] doublet, an
identification consistent with the source being the object in the 
(LCIRS) B band image, located less than $1\farcs5$ from our estimated position.
The only other single-line object with a B-band candidate this close to
our position is \m89-2, which we also exclude from the candidate LAEs 
on the basis of a probable B-band detection.

%
All the single-line candidates
show at least a faint smudge in the WB48-78 image within 1\farcs5 of 
our position.  If these faint objects are the counterparts of our
LAEs, they should be brighter in the z band. These galaxies are not 
detected in the LCIRS-z images, but the depth is not sufficient to 
rule out a continuum break across the line.  It is only the high 
frequency of plausible [OII] interlopers in the 10H field that
strongly suggests many will turn out to be foreground interlopers. 
The extreme depth of the WB48-78 image does demonstrate that
plausible continuum counterparts can be found at offsets from
the emission-line objects that are significantly smaller than
those from bright objects with photometric-redshifts in the 10H 
field.  This result strengthens our suspicion that the objects
near redshift 1.2 that are several arcseconds from our emission-line 
position are {\it not} the same source, although the two objects
may be associated (i.e. at the same distance).

\subsection{Completeness and Effective Area}  \label{sec:complete}


To determine completeness levels, we added objects of various 
brightness to the search frames and then checked whether SExtractor
recovered them.  The intensity of an object was modeled by a 
 gaussian profile ranging in size from 0\farcs45 to
0\farcs90 spatially and linewidth 6\AA\ to 12\AA. 
We tuned SExtractor to find most of our by-eye sources but not to go
too much fainter, which yields many spurious sources. This tuning
requires that an object have at least 10 pixels above 1.5 times the
background rms. SExtractor found ~90\% of our by-eye sources and
most of the 'missing' 10\% can be explained as being near image
artifacts that confuse SEextractor. This tuning of SExtractor
allowed us to recover the simulated
sources in a way that was both automated and very close to 
what the human eye finds.

\begin{figure}[h]
      {\includegraphics[scale=0.6,angle=-90,clip=true]{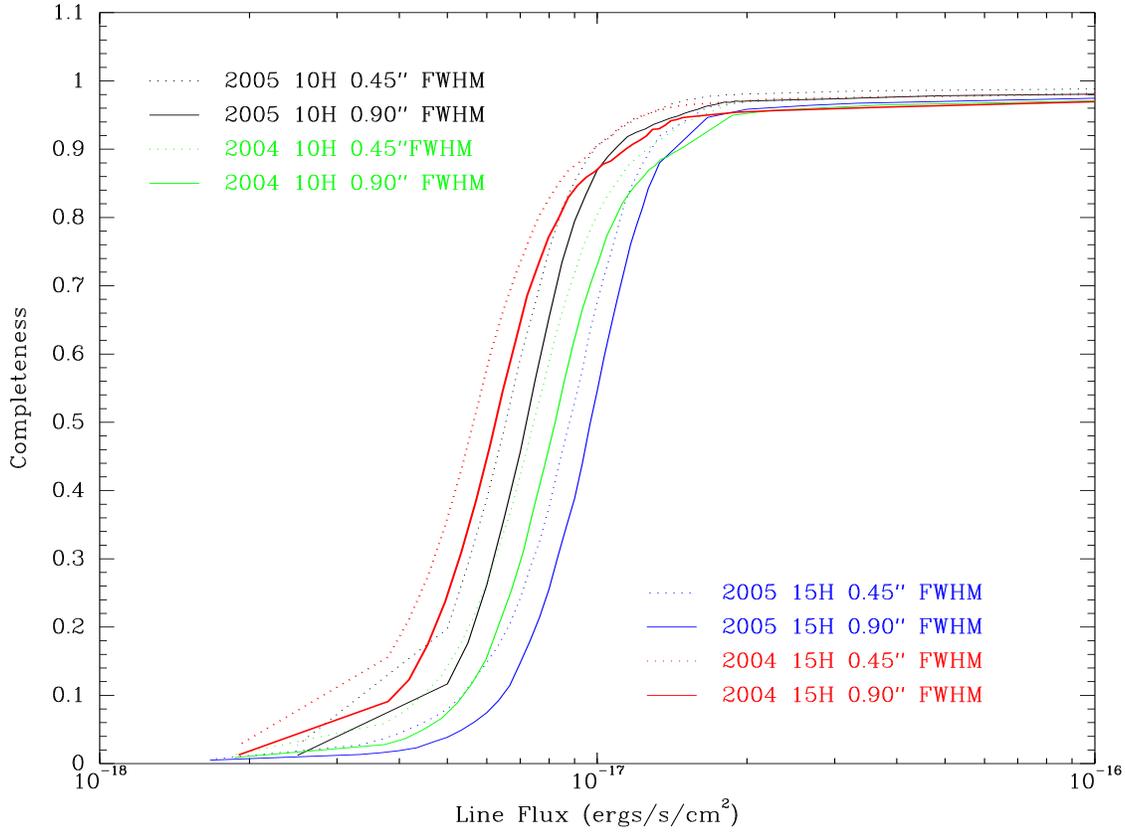} 
        }
          \caption{Completeness of emission-line surveys for
	    two limiting sources sizes. For the 2005 search in
	    the 10H Field, the line flux corresponding to 50\%
	    completeness is $6.6 - 7.2 \times 10^{-18}$\flux\
	    for sources between 0\farcs45 and 0\farcs90 FWHM.
          }
           \label{fig:c} \end{figure}

Results of the Monte Carlo simulations are shown in
\fig~\ref{fig:c}. Most of our sources were unresolved spatially, 
so we interpolate to the results for 0\farcs7 seeing.
The recovery rate of our survey is 50\%
at a line flux of  $6 \times 10^{-18}$\flux\ for the most
sensitive search, the 2004 observations of the 15H field, 
degrading to $9 \times 10^{-18}$\flux\ for the least
sensitive search, the 2005 15H field. 

\begin{figure}[h]
     {\includegraphics[scale=0.5,angle=0,clip=true]{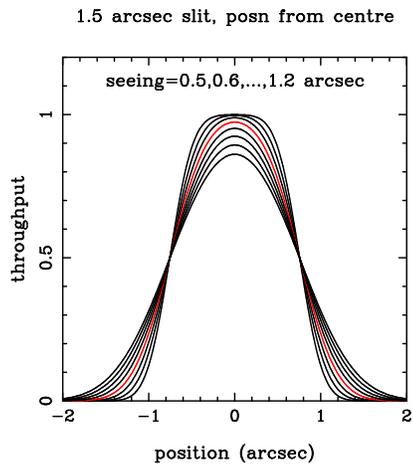} 
        }
          \caption{Attenuation of flux vs. offset from center of 1\farcs5
	    wide slit. Slit losses are shown for seeing-convovled sources 
	    with Gaussian surface brightness profiles of FWHM 0.5 to
	    1\farcs2 in steps from 0\farcs1. The red curve represents the
	    worst effective image quality in our survey.
          }
           \label{fig:slit} \end{figure}

Attenuation by the slit and masked areas such as chip edges, 
the zero-order image of the mask, and bright foreground objects 
affect the effective area of our survey. The Monte Carlo simulations 
quantify how the  slit length that we can search is reduced
by masked objects.  For a fixed slit width of 1\farcs5,  this effective 
slit length per IMACS pointing ranged from 45.134$\square ^{\arcmin}$  in 
the 15H 2005 search to 45.589 $\square ^{\arcmin}$ in the 10H 2005 search. 
The effective area over which a particular source could have been
detected  depends on its brightness. Emission-line objects near our 
detection threshold will only be detected if a slit is centered on
them, while extremely bright sources could be highly attenuated and
still detected.  These slit losses were modeled using a Gaussian 
brightness profile as illustrated in Figure~\ref{fig:slit}. 

We modeled the effective slit length and completeness separately
for each of our four search runs. The calculations in the next section 
use the results.


\section{The LAE Galaxy Population}

In this section, we discuss the properties of the
LAEs, compute their space density, and find the
families of Schechter function parameters that can
describe  this galaxy population. These descriptions
of the LAE population are used to estimate the
production rate of Lyman continuum photons, which
is compared to the ionization requirement of the IGM 
at redshift 5.7.

\subsection{Properties of LAEs} \label{sec:properties}

The \lya\ fluxes that we measured correspond to luminosities
of $\log \lya ({\rm erg~ s}^{-1}) = 42.70$, 42.81, and 42.89 
for \m80+3, \m29.5+5, and \m71-5, respectively.  These LAEs are 
fainter than the luminosity break (i.e. $L_*$) fitted to a
larger sample of LAEs in the Subaru Deep Field (Shimasaku \et
2006). Other surveys that have confirmed LAEs this faint at
$z \sim 5.7$ include:  GRAPES -- the  Grism ACS Program for 
Extragalactic Science (Pirzkal \et 2007) in the Hubble Ultra
Deep Field (HUDF, Beckwith \et 2006), the Gemini Deep 
Deep Survey (GDDS) -- which obtained 30~hr spectrosopic observations  
of  i-dropouts in the HUDF (Stanway \et 2006), a survey of 
cluster caustics for lensed LAEs (Ellis \et 2001), and the
narrowband imaging  surveys of Hu \et (2004) and
Shimasaku \et (2006) with Subaru.

We estimate star formation rates for the Magellan LAEs 
from their line luminosities.  Carrying the amount of 
\lya\ attenuation, $f_{lya}$, along as a parameter,
the Kennicutt (1998)  conversion from \Ha\ luminosity 
to SFR and the \lya to \Ha flux ratio from Case B 
recombination (Brocklehurst 1971) yield star-formation
rate estimates
\begin{eqnarray}
SFR = 1.0 f_{Ly\alpha}^{-1} \msunyr (L_{Ly\alpha} / 1.05 \times 10^{42} 
{\rm ~ergs~s}^{-1}). 
\end{eqnarray}
This expression assumes continuous star formation, solar
metallicity, and a stellar initial mass function with 
dlogN/dlogM = -2.35 from 0.1 to 100\msun. 
The estimated SFRs of the LAEs are $4.8f_{lya}^{-1}$, 
$6.1f_{lya}^{-1}$,  and $7.4f_{lya}^{-1}$\msunyr.
Variations in the ratio $ L(Ly\alpha)/L_{\nu}(UV)$ are
large among LAEs, but star formation rates derived from 
the UV continuum are typically a factor of a few times higher 
than those estimated from \lya (e.g. Pirzkal \et 2007).  
The faint LAE population is very blue 
(Stanway \et 2006; Pirzkal \et 2007), so the upward 
corrections to these UV luminosities for extinction may
be small.  Our guess for the new LAEs is  an attenuation 
$f_{lya} \sim 0.25 - 0.5$, which corresponds to SFRs of 
$\sim 10-30 $\msunyr.

%

%

Broadband photometry of the COSMOS field
(and the LCIRS field) is not currently deep enough to
detect the LAEs from the Magellan MNS  survey. Rest-frame 
UV-to-optical colors are available for nine LAEs in the 
HUDF at slightly lower redshift $4.0 < z < 5.7$. Since the
\lya\ properties of the two samples are similar, the stellar 
masses estimated by Pirzkal \et (2007) for the former, 
$10^6$ to $10^7$\msun, may be indicative of the masses of
the Magellan MNS LAEs. The LAEs are not resolved in good 
seeing, so the size of the emitting region is less than 
about 5~kpc. 


The shapes of the \lya line profiles contain additional information 
about the properties of LAEs and the intervening IGM. Due to the high 
scattering cross section of the \lya\ transition,  modeling
the diversity of \lya\ profiles observed in star forming
galaxies requires radiative transfer calculations (Ahn \et 2003;
Hansen \& Oh 2006; Verhamme, Schaerer, and Maaselli 2006; Tapken \et 2007).
Outflows are thought to decrease the
intrinsic \lya\ attenuation via frequency shifting, so their
presence may make it easier to detect \lya emission.
Since the Magellan LAEs have high SFRs, we interpret the
line profiles in terms of the properties of galactic winds 
blown by the LAEs.

We measure \lya\ linewidths of 260\kms and 240\kms FWHM, 
respectively, for \m80+3 and \m29.5+5. The \lya\ line 
in the Keck spectrum of \m71-5 is only 98\kms wide, but
the line is broader in the low-resolution Magellan spectrum.
We attribute the discrepancy to spatial variations across
the object, which are weighted differently by seeing and 
slit alignment.\footnote{ 
       The same object is detected by both observations
       because independent wavelength calibrations put 
       the line at the same wavelength in both spectra.}
We fitted very simple single-component and two-component models to 
the  high-resolution line profiles.  The one component model is a Gaussian 
with all flux blueward of the peak set to zero to mimic absorption
by intergalactic gas. This kernel was convolved with another Gaussian 
that represents the instrumental broadening.   The 2-component model 
has an additional, weaker Gaussian added redward of the main line to  
simulate backscattering off  an outflow.

The lower right panels of Figures~\ref{fig:msdm29.5+5}, 
\ref{fig:msdm80+3}, and \ref{fig:msdm71-5} show the results.
The profile of \m29.5+5 is well fitted with a single component.
The fitted instrumental smoothing, $\sim 110$\kms FWHM, is consistent 
with the resolution of the Keck-LRIS, follow-up spectrum. 
The model indicates a FWHM prior to attenuation of 400\kms. 
The high-resolution spectrum of \m80+3 shows a
red wing. For an instrumental resolution of 95\kms, the
fitted single-component model has a width of 300\kms prior
to attenuation. A spectrum with higher signal-to-noise 
ratio is needed to determine the reality of additional
line components, such as a possible narrow component located
300\kms redward of the main line.

%

We fitted a model introduced by Dawson \et (2002) and 
Hu \et (2004) that is now in common use. 
Westra \et (2005, 2006) fitted the model to two extremely luminous
LAEs at $z \sim 5.7$ from WFILAS. The two-component model for \m80+3 
reminds one of 
their result for LAE S11\_5236, which has a second peak
offset 400\kms redward of \lya that is only 20-90\kms wide.
This model has also been fitted to the composite line profile
 of LAEs discovered serendipitously in the DEEP2 redshift 
survey (Sawicki \et 2007).   The intrinsic width of the primary component 
$\sim 350$\kms, and the component fitted to the red wing is offset 
$\sim 420$\kms.  These values are similar to what we find for the 
Magellan LAEs even though the DEEP2 LAEs are more luminous (mean 
$L_{Ly\alpha} = 6 \times 10^{42}$~erg~s$^{-1}$) and found at lower 
redshift (median redshift 4.1). The substantial star formation rates 
suggest that winds are the likely source of the redshifted emission,
but gas inflow can produce similar line profiles (Dijkstra \et 2006).
Strong winds would play a critical role facilitating the 
escape of ionizing photons (Fujita \et 2003).



\subsection{Number of LAEs Detected} \label{sec:counts}

We confirm three LAEs from the four survey masks. The total 
unobscured slit area and the filter bandpass at FWHM sample a 
volume of $4.50 \times 10^4 h_{70}^{-3}$~Mpc$^{3}$. The slit
and the filter attenuate every LAE.  The amount of attenuation
is unknown for any individual source but is well understood
statistically for the population. For any model of the luminosity
function, we can accurately predict the attenuation of objects
drawn from such a population and compute the number of sources
detected given the completeness limit of each of our four
search observations.


For parameterizations of the luminosity function 
($L_{*}$, $\phi_*$, $\alpha$), where 
\begin{eqnarray}
d \phi(L) = \phi_* (L/L_*)^{\alpha} e^{-L/L_*} d(L/L_*),
\end{eqnarray}
we computed the mean number of LAEs that our survey would detect. We calculated
the volume from which a source is drawn using the slit width and redshift interval 
over which the  source could have been detected. Because we start with the
intrinsic luminosity (or flux), we can model the mean attenuation precisely.
We discuss the results obtained using our 90\% completeness level.

\begin{figure}[h]
     {\includegraphics[scale=0.7,angle=0,clip=true]{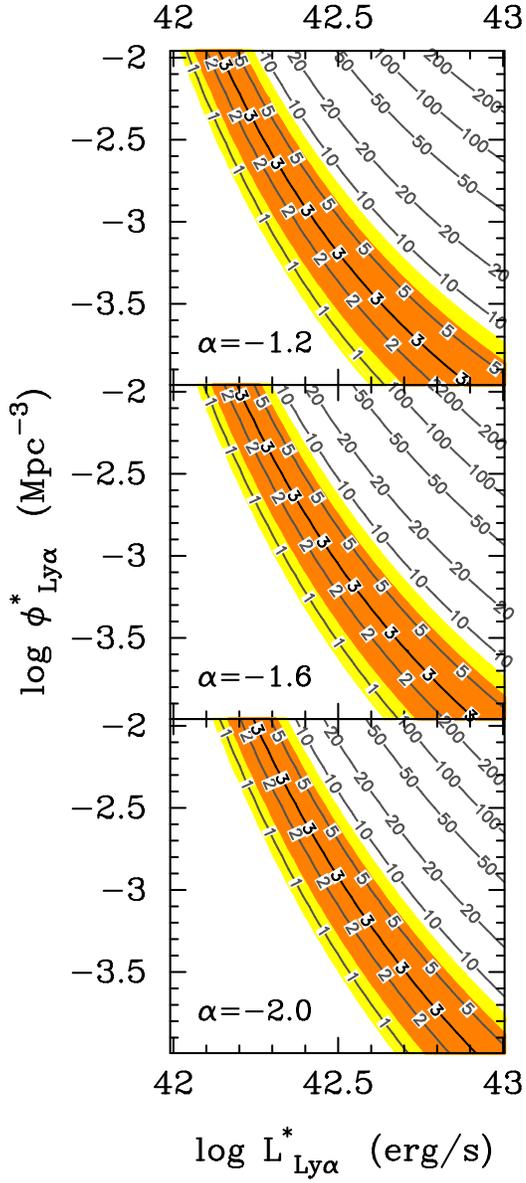}
        }
        \caption{Mean number of LAEs detected in our experiment for different
	  values of the intrinsic luminosity function parameters. From top
	  to bottom, the panels show faint-end slope $\alpha = -1.2, 1-.6, {\rm ~and} 
	  -2.0$. The y-axis is the normalization of the luminosity function, $\phi^{*}$,
	  The x-axis is the characteristic luminosity $L_{*}$.
	  The orange and yellow shading marks the 84.1\% and 95\% confidence regions 
	  for dectecting 3 LAEs.
          }
           \label{fig:Nmodel} \end{figure}

Figure~\ref{fig:Nmodel} illustrates the range of luminosity
function parameters that yield $\sim 3$ detections.
The shaded regions in \fig~\ref{fig:Nmodel} denote the allowed 
combinations  of $(\phi_{*}, L_{*})$ by requiring that the model yield 1.37 to 
5.92 LAEs (85\% confidence limits, orange) or $3^{+4.75}_{-2.18}$ LAEs (95\% CL
in yellow), where the limits were computed from Poisson statistics (Gehrels 1986). 
The number density of LAEs in the Magellan MNS survey constrains 
the product $L_* \phi_*$ well.
The contours of equal number show how lower normalization values, $\phi_*$,
require higher $L_{*}$ to yield a fixed number of objects $N$. 
 Since our data do not constrain the faint-end slope well,
we show results for three assumptions $\alpha  = -1.2, -1.6, {\rm ~and} -2.0$.
 In the limit $\alpha = -2.0$,  a normalization similar to
that at $z \sim 4.9$ in the SDF, $\phi_* \approx 0.0055$~Mpc$^{-3}$ (Ouchi \et 2003)
requires $L_*$ lie in the range $1.78 \times 10^{42}$ to $5.97 \times 10^{42}$\ergsec.
As the faint-end slope becomes flatter, the characteristic luminosity must 
increase to yield 3 detected objects if the normalization is held fixed.  


Based on our confirmation rate to date, additional spectroscopic follow-up will 
confirm, at most, a couple more LAEs in our emission-line survey.  The number of 
remaining line sources for which a LAE identity has not been ruled out provides
a strict upper limit.  Our follow-up observations were remarkably
complete; only 4 objects in the 10H field, and 7 candidates in the 
15H field, were not re-observed without the blocking filter.
Some emission-line objects, however,  were not-recovered in follow-up 
spectroscopy (8 and 7 galaxies in the 10H and 15H 
fields, respectively). In the previous section, we argued that only two 
single-line objects in the 10H field were not likely foreground emitters. 
To compute
an upper limit, however, we count all 9 single-line sources (including 
the confirmed LAE \m80+3).  We found 7 single-line objects including
2 confirmed LAEs in the 15H field.  The total number of remaining
LAE candidates is 42.  For comparison, adopting the Shimasaku \et luminosity 
functions with $\alpha = -2.0, -1.5, {\rm and~} -1.0$ yields 17, 14, and 12 LAEs,
respectively, in our IMACS-MNS survey. That our upper limit is well above this 
confirms that most of our emission line objects that remain unidentified 
are not LAEs as we have argued.



\subsection{Cumulative LAE Luminosity Distribution}


Monte Carlo simulations of the luminosity distribution are
the preferred method for determining the best luminosity
function parameters.  For a larger (or deeper) survey with 
more sources than our survey, we would estimate the luminosity 
function parameters by drawing sources randomly  from a model 
of the luminosity function, assigning them random positions in the 
field and applying the detection biases associated with the 
slit, filter, and data processing unique to MNS.  The likelihood 
of the true parameters lying in any particular range could then be 
determined. 

With just three confirmed LAEs, however, we use a simple variant 
of the  $1/V_{max}$ method (Felten 1976; Dawson \et 2007) to
estimate the cumulative luminosity distribution.   We model
the maximum volume, $V_{max,i}$, within which each source could have 
been recovered. We use the slit width and redshift interval within 
which the object would have a flux greater than that of
our survey's 90\% completeness limit. Due to the high fraction of 
candidates followed up and recovered, the correction for 
unobserved candidates is small and omitted.
Proceeding from the brightest source to the faintest, we add up 
the number density of LAEs to obtain the cumulative luminosity 
distribution. Our best estimate is a lower bound in some sense
because of the possible attenuation of each LAE. 
Table~\ref{tab:cumlf} shows this cumulative  luminosity function.

\begin{figure}[h]
     {\includegraphics[scale=0.5,angle=-90,clip=true]{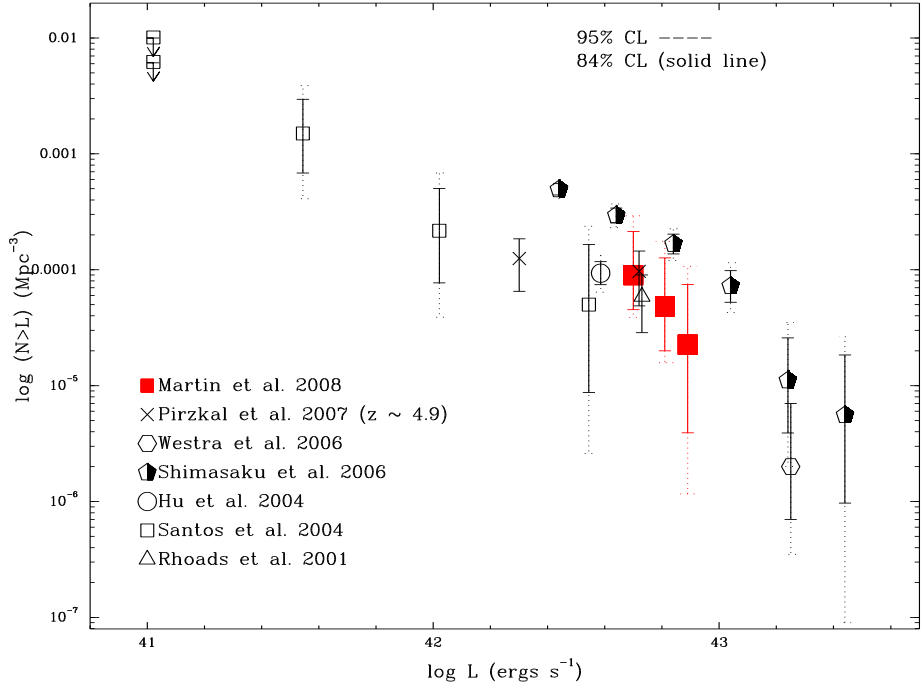} 
        }
        \caption{Cumulative number density of LAEs. Our best
	  estimate is shown by the solid squares; the errorbars are
	  discussed in the caption to Table~\ref{tab:cumlf}.
	  We compare number densities from the following narrowband 
	  imaging surveys:  LALA 2001 (triangle), Hu \et 2004 (circle), 
	  Westra \et 2006 (hexagon), and  Shimasaku \et 2006 (pentagons). 
	  Spectroscopic results are shown from GRAPES
	  (Pirzkal \et 2007, crosse) and a program that
	  positioned longslits along cluster caustics
	  (Santos \et 2004, open squares). 
          }
           \label{fig:cumlf} \end{figure}

\fig~\ref{fig:cumlf} compares this distribution (solid squares) to 
counts from other surveys at redshift 5.7. The open squares denote
faint LAEs  from Santos \et (2004) that are gravitationally lensed 
by a foreground galaxy cluster. This ground-breaking survey of cluster
caustics demonstrates the existence of less luminous LAEs; but the LAE
number density is far more uncertain than the statistical error (shown).
The survey volume is difficult to estimate and very small (i.e.
subject to cosmic variance), so one may not wish to weight these points 
heavily when constraining the luminosity function. The GRAPES produced
three LAEs at $ z > 5$;  the density from Pirzkal \et (2007), which
includes lower redshift LAEs, is denoted by a cross in \fig~\ref{fig:cumlf}.
The luminosities and redshifts of the LAEs discovered by Hu \et
(2004, open circles) with wide-field narrowband imaging provide the most 
direct comparison to our Magellan survey. The latter, like many of the imaging 
surveys, was carried out with the SuprimeCam mosaic CCD camera on Subaru but 
is unique among them; a high percentage of candidates have been 
confirmed spectroscopically. The number density of LAEs estimated from our 
Magellan MNS survey agrees very well with the result from the 
survey of  Hu \et, which sampled a volume comparable to Shimasaku \et

Number density estimates from the Large Area Lyman-Alpha survey
still require significant extrapolation for unconfirmed LAEs at 
redshift 5.7 (Rhoads \et 2001). Their best estimate (Malhotra \& 
Rhoads 2004, open triangle) is, however, consistent with our new result.
The data at redshift 5.7 will improve with the spectroscopic confirmation 
of other large surveys with SuprimeCam  (Murayama \et 2007;
Ajiki \et 2003, 2004, 2006). In Figure~\ref{fig:cumlf}, we show 
LAEs, selected photometrically from SuprimeCam imaging of the Subaru Deep 
Field (SDF); Shimasaku \et (2006) published spectroscopy 
for about one-third of the sample. 
For  a faint-end powerlaw slope of  $\alpha = -2.0$, their fitted
characteristic luminosity is $L_{*} = 1.6^{+0.9}_{-0.6} \times 
10^{43}$\ergsec\ with normalization $\phi^{*} = 1.6^{+1.4}_{-0.7} 
\times 10^{-4}$~Mpc$^{-3}$;  and  the integral of this 
luminosity function is plotted in \fig~\ref{fig:cumlf_fit}. 
The LAE number density of the Shimasaku \et LAEs is 3-6 times higher 
than our result over the luminosity range $\log L = 42.8^{+0.1}_{-0.1}$. 
Inspection of the error bars in \fig~\ref{fig:cumlf} shows that only
at the low end of this range are our results even marginally consistent.

It is not obvious to us that cosmic variance is the explanation of this 
discrepancy. Stark, Loeb, \& Ellis (2007) argue quantitatively that the variance in 
the Subaru Deep Field LAE Survey at redshift 5.7 is dominated by Poisson
shot noise rather than cosmic variance. The area imaged, about 0.20 square 
degrees, is similar to that covered by both the LALA survey and Hu \et (2004).
The shallow narrowband imaging survey, WFILAS (Westra \et 2006),  covers a larger 
area. Only two of 7 LAE candidates have been confirmed in their 0.72 sq. degree 
field, but these two objects are very luminous; and we show their number 
density in \fig~\ref{fig:cumlf} (hexagon).  To estimate the total variance 
in our survey, we used the on-line calculator described by Trenti \& 
Stiavelli (2008). To simulate the effect of the sparse sampling of the
slits, we used a low value for the survey completeness (10\%) and
the combined survey areas in the 10H and 15H fields.  For any halo filling 
fraction less than 70\%, the total variance was dominated by the Poisson
term rather than the clustering term. We conclude that cosmic variance is
unlikely to dominate the uncertainty in our measurement of the LAE number
density.

The Shimasaku \et result in the SDF is consistent with the WFILAS result
considering their uncertainties, but the estimated values differ by
a factor of five. 
Since several of these surveys probed large volumes, we 
see no reason to favor the newer Shimasaku \et result until
spectroscopic confirmation is completed.
The combined data do not 
constrain the break luminosity, $L_{*}$, well in \fig~\ref{fig:cumlf}.




\subsection{The \lya Escape Fraction} \label{sec:uv}

\begin{figure}[h]
     {\includegraphics[scale=0.5,angle=-90,clip=true]{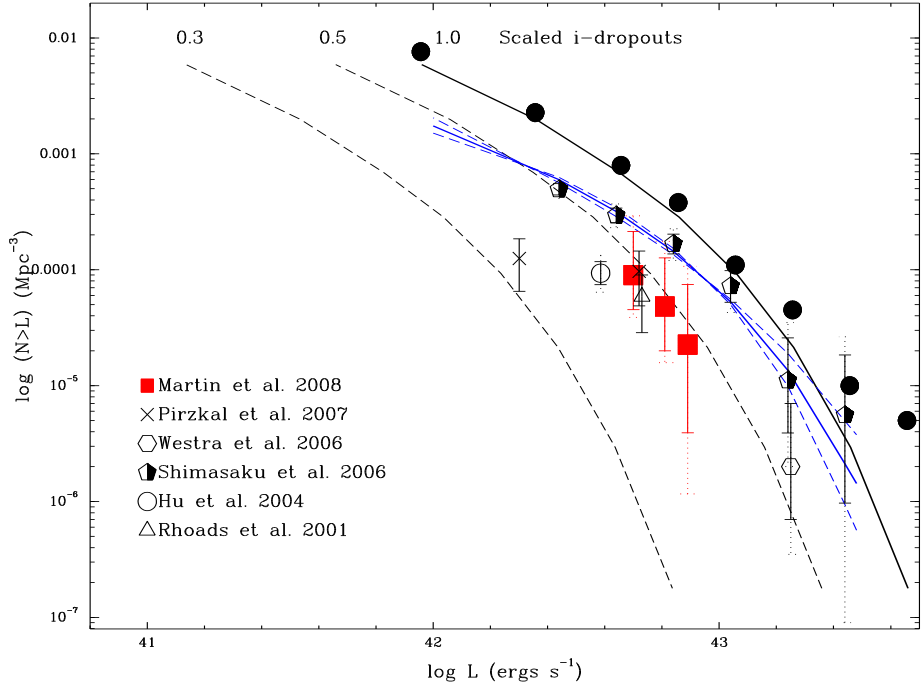} 
        }
        \caption{Comparison of cumulative number density of $z=5.7$
	  LAEs and i-dropouts.  Blue curves show the fitted luminosity
	  distribution of LAEs from Shimasaku \et (2006) with faint-end
	  slope $\alpha = -2.0, -1.5 (solid), {\rm ~and~} -1.0$. Black curves
	show the adopted fit to the distribution of i-dropouts from 
	Bouwens \et (2007); for purposes of illustration, we adopt
	 constant \lya\ escape fractions of $f_{lya} = 1.0, 0.5, {\rm ~and~} 
	 0.3$. The dashed, black curve illustrates the magnitude of
	 some systematic uncertainties in the luminosity distribution fit;
	 and solid circles show the i-dropout counts, converted to
	 \lya\ luminosity using $f_{lya} = 1.0$. 
          }
           \label{fig:cumlf_fit} \end{figure}

The luminosity distribution of i-dropouts has been measured to
very faint limits in the HUDF and a few other fields with deep
HST/ACS observations (Bouwens \et 2007; Bouwens \et 2006). Since 
we do not detect continuum from our faintest LAEs, and spectra have not
been acquired for all of the i-dropouts, it remains unclear whether
the \lya -selected sample is drawn from the i-dropout population. 
For purposes of illustration, we mapped the rest-frame UV luminosities,
 $L_{\nu}$, of objects drawn from the i-dropout luminosity function into 
line emission. We used the relation
$L(\lya) = 1.47 \times 10^{42} {\rm ~erg~ s}^{-1} f_{Ly\alpha} 
(L_{\nu} / 10^{28} {\rm erg~s}^{-1}~{\rm Hz}^{-1}$),
which assumes a timescale for star formation inherent to the 
mapping from UV luminosity to SFR given by Kennicutt (1998).
\fig~\ref{fig:cumlf_fit} shows the cumulative \lya\ luminosity 
function  of the i-dropouts.  The three models are labeled 
by the \lya\ attenuation, $f_{Lya}$.  For the no attenution
example, this heuristic model overpredicts the number density of 
LAEs.



The surprise is how easy it is to reconcile the number density of LAEs 
and i-dropouts. Absorption by intergalactic gas significantly attenuates the
\lya\ emission from $z > 4$ galaxies.  For  redshift $\sim 6$
galaxies, the SFR inferred from the UV luminosity is typically several
times higher than that inferred from the \lya\ luminosity.
The dashed curves in \fig~\ref{fig:cumlf_fit} show that absorbing 
the blue half of the line profile, equivalent to setting $f_{Ly\alpha} 
= 0.5$, is one way to produce the LAE luminosity distribution
from the i-dropout population. This particular model is not unique
and may be a strong oversimplification of the relation between these
two galaxy populations.   The ratio of UV to \lya\ 
luminosity presents enormous scatter among galaxies. The escape fraction
$f_{Lya}$ could vary strongly with luminosity (or halo mass).

Additionally, if the fraction of galaxies 
presenting \lya\ emission is less than unity, then the i-dropout 
luminosity function would need to be shifted in number besides 
star formation rate in order to model the LAE luminosity function.   
We know that only 25\% of redshift 3 LBGs are 
LAEs (Shapley \et 2003). The data for $z \sim 6$ galaxy samples
appear to disagree.  Shimasaku \et (2006) and Kashikawa \et (2006) 
argue for a rise in the fraction of galaxies with \lya emission
to near unity at $z \sim 6$, while Dow-Hygelund
\et (2007) found many i-dropout galaxies without \lya emission.
Modeling the effect of color cuts and equivalent width cuts on these 
samples will be required to clarify the \lya emission fraction.
The correct physical mapping between the i-dropout and LAE populations 
is not currently known. Our comparison simply shows that a simple 
model with 50\% \lya\ attenuation
and a high fraction of LAEs among i-dropout galaxies is
consistent with our estimated density of LAEs. 
That it severely overpredicts the counts from the lensed 
observations could mean a number of things: smaller \lya\
escape fractions from dwarfs, an overestimate of the number 
of faint i-dropouts, or an underestimate of the number
density of lensed LAEs.




\subsection{Reionization Constraints} 



The presence of Gunn-Peterson troughs in the spectra of $z \sim 6$ quasars 
(Becker \et 2001; Djorgovski \et 2001) requires only a tiny neutral gas 
fraction in the IGM, about one in $10^4$ H atoms.  The absence of strong
evolution in the shape of the LAE luminosity function between redshift
5.7 and 6.5 suggests little change in the neutral fraction leading many to 
believe reionization was complete by redshift 6.5 (Malhotra \& Rhoads 2004; cf.
Kashikawa \et 2006). A protracted period of reionization from redshift 
14 to 7 would be consistent with
the Thompson optical depth inferred from measurements of the cosmic microwave 
background (Spergel \et 2007).  The time from redshift 7 to 5.7 is 229~Myr,
long enough that the intergalactic gas would recombine were it not bathed
with Lyman continuum radiation.\footnote{The recombination time depends
                on the clumping factor of the IGM. Following 
                Madau \et (1995), $t_{Rec} = 290~Myr~ C_{6}^{-1}
                \left( \frac{6.7}{1+z}  \right)^3 \left( 
		\frac{0.047}{\Omega_b h_{70}^2} \right)$.}
We therefore ask whether the LAE galaxy population discovered to date might
supply these ionizing photons.

To maintain IGM ionization at $z = 5.7$, the number of
ionizing photons escaping from galaxies over the recombination timescale 
must equal or exceed the number of hydrogen atoms.  The recombination rate is
faster in denser regions of the IGM, so the required number of escaping 
photons increases with the clumpiness of the gas. The gas is smoothest on 
large scales and clumpier in the virialized regions where galaxies form.
The gas in these halos absorbs a high fraction of the ionizing
photons (produced by galaxies), so the Lyman continuum escape fraction,
$f_{LyC}$, is lower on the larger scale of the IGM.  For the regions
of IGM penetrated by ionizing radiation, a clumping factor can be calculated 
that is the reciprocal of the volume filling factor of protons (or electrons).
The clumping factor is defined here in units of $C \equiv  <n_{HII}^2> / 
\bar{n}_{HII}^2$ (Madau \et 1999); and at $z \sim 6$ it is likely between 
1 and 10 (Furlanetto \et 2006; Gnedin 2007). We parameterize the clumping 
factor in units of $C = 6$ such that $C_{6} \equiv ( 1 / 6) <n_{HII}^2> 
/ \bar{n}_{HII}^2$. Following
Madau \et (1999), the critical {\it ionization rate in the IGM} is 
$\dot{N}_H = \bar{n}_H(0) / \bar{t}_{rec}(z)$ per unit comoving 
volume, which we can write as
\begin{equation}
\dot{N}_{H} = 5.2 \times 10^{50} {\rm ~s}^{-1}
{\rm ~Mpc}^{-3}~ C_{6} \left( \frac{1 + z}{ 6.7} \right)^3
\left(  \frac{\Omega_b h_{70}^2}{0.047} \right)^2 .
\label {eqn:ncrit} \end{equation}

The required  {\it production rate of H ionizing photons in galaxies}
is much higher because little Lyman continuum radiation escapes 
(Hurwitz \et 1997; Leitherer \et 1995).  Measurements at redshift 
3 suggest absolute escape fractions $f_{LyC} \sles\ 0.1$ 
(Chen \et 2007; Shapley \et 2006; Inoue \et 2005; Fernandez-Soto 
\et 2003; Giallongo \et 2002; Steidel \et 2001). Numerical simulations 
suggest escape from the virialized halos of low mass galaxies is
difficult (Gnedin \et 2007) and that the result is sensitive
to the interplay between the blowout of a galactic wind and 
the star formation history (Fujita \et 2003).
In the absence of any contribution from quasars, the minimum cosmic
star formation rate\footnote{ 
  Here, we have used an initial stellar mass
  function $dN/dM \propto M^{-2.35}$ from 1.0\msun\ to 100\msun. 
  Continuous star formation at a rate of  1.0\msunyr\ produces ionizing
  photons at a rate of $\dot{N}_H = 10^{53.4}$~s$^{-1}$ (Kennicutt 1998;
  Leitherer \et 1999).}
required to keep the universe ionized at z=5.7 is 
\begin{eqnarray}
 \dot{\rho}_{*} = 0.02 \msunyr {\rm Mpc}^{-3} \times \nonumber \\
~C_{6} f_{LyC,0.1}^{-1} 
\left( \frac{1 + z}{6.7} \right)^3 \left( \frac{\Omega_b h_{70}^2}{0.047} \right)^2,
\end{eqnarray}
where $f_{LyC}$ is the fraction of Lyman continuum photons escaping from
galaxies, and $f_{LyC,0.1}$ means  $(f_{LyC} / 0.1)$.
 The ratio $C /  f_{LyC}$ is not very sensitive to spatial scale,
so the spatial dependencies of the clumping factor and escape fraction 
cancel each other out to some extent.

To determine what fraction of this ionization requirement
the source population provides, we can add up all the LAEs,
weighting by luminosity, in a given volume of space. Using
Case~B recombination conditions and Equation~\ref{eqn:ncrit},
we find the critical \lya\ luminosity density required for
full IGM ionization is
\begin{eqnarray} \label{eqn:lden}
\L_{Ly\alpha}  = 3.0 \times 10^{40} {\rm ~erg~s}^{-1} {\rm Mpc}^{-3}~ 
\times \nonumber \\
C_{6} (1 - 0.1 f_{LyC,0.1}) \left( \frac{f_{Ly\alpha,0.5}} {f_{LyC,0.1}} \right)
\left( \frac{1 + z}{6.7} \right)^3 \left( \frac{\Omega_b h_{70}^2}{0.047} \right)^2 .
\label{eqn:lumdensity} \end{eqnarray}
This equation highlights some obvious facts.
If \lya radiation is highly attenuated (i.e. $f_{Ly\alpha} << 1$), 
then the observed \lya\ luminosity density can be low even though the
observed galaxy population ionizes the IGM. Likewise, if Lyman continuum 
leakage from galaxies is very low  (i.e. $f_{LyC} << 1$), then 
even a high number density of galaxies will not completely ionize
the IGM.  To visualize the implications of the uncertain gas physics
for the IGM ionization requirement, we collect all the factors in 
Equation~\ref{eqn:lumdensity} describing the gas physics into a single 
parameter 
\begin{eqnarray}
\zeta = C_{6} (1 - 0.1 f_{LyC,0.1})  f_{Ly\alpha,0.5} 
f_{LyC,0.1}^{-1}, 
\end{eqnarray}
which is likely between 0.1 and 2. Although
our simple modeling does not include possible variations in $\zeta$
among galaxies, we note that our survey limit is brighter than 
that expected at the theoretical mass cut-off for escape of
Lyman continuum photons (Gnedin \et 2007; Gnedin 2007).

\begin{figure}[h]
     {\includegraphics[scale=0.7,angle=0,clip=true]{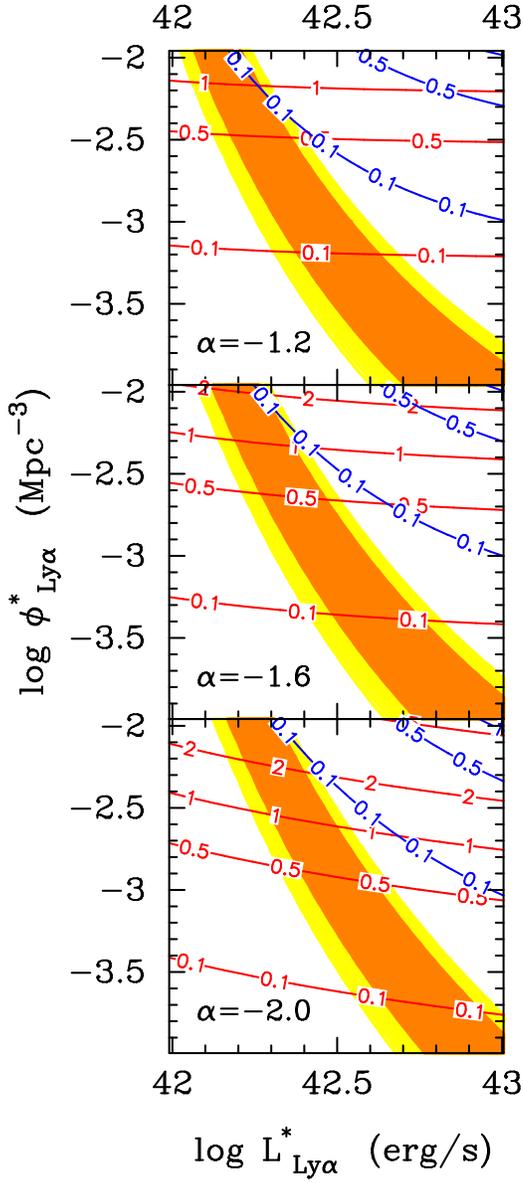}
        }
        \caption{Total \lya luminosity density for the 3 values of the
	  faint-end slope. Two lower limits are illustrated:  (1) our survey 
	  detection limit of $\log L = 42.57$~erg~s$^{-1}$ (blue contours), 
	  and (2) an extrapolation
	  to $\log L = 41.00$~erg~s$^{-1}$ (red contours), 
	  which corresponds to dwarf galaxies with $SFR = 
	  0.095 f_{Ly\alpha}^{-1}$.  The critical number of sources required
	  to keep the IGM ionized is marked in units of the parameters
	  $C_{6} (1 - 0.1 f_{LyC,0.1}) f_{Lya\alpha,0.5} / f_{LyC,0.1} = 1.0$.  
	  With the lower limit
	  in (1), it appears unlikely that enough galaxies have been discovered 
	  to keep the IGM ionized at z=5.7. It appears, however, from (2) that
	  an extension of this population to lower luminosities, even with a shallow 
	  faint-end slope like $\alpha = -1.2$, could easily produce the missing
	  ionizing photons.
          }
           \label{fig:Lmodel} \end{figure}

We consider the luminosity density of the LAE population for two
minimum luminosities: the Magellan MNS detection limit of 
$\log L_{Ly\alpha} ({\rm erg~s}^{-1}) = 42.57$  and an 
extrapolation to $\log L_{Ly\alpha} ({\rm erg~s}^{-1}) = 41.0 $.  
This lower value corresponds to a $SFR \approx 0.1$\msunyr, a level
of activity common among dwarf starburst galaxies at low redshift.
The line flux of such objects, $F_{Ly\alpha} \approx 3 \times 10^{-19}$\flux\ 
at redshift 5.7, is well below current detection limits.
For each luminosity function represented by the grid of Schechter 
function parameters in \fig~\ref{fig:Nmodel}, we integrated the 
\lya\ luminosity density down to the two minimum luminosities.
The total luminosity densities are shown in \fig~\ref{fig:Lmodel}.  
Only those values within the (orange and yellow) shaded region can be
attained by a LAE population consistent with our number counts.

The \lya\ luminosity density in \fig~\ref{fig:Lmodel} is shown
relative to that required for IGM ionization by Eqn.~\ref{eqn:lden}.
Contours of equal luminosity density are labeled by the value of 
$\zeta $ required for ionization; blue and red contours correspond
to a minimum luminosity $\min \log L_{Ly\alpha}(erg/s) = 42.57$ 
and 41.0, respectively, for the lower integration limit.  If spatial 
variations in  $C_{6}$ and 
$f_{LyC,0.1}$ do largely cancel out, then $\zeta$\ largely 
reflects the amount of \lya\ attenuation, $f_{Ly\alpha,0.5}$.
 We consider values of $\zeta$ from 0.1 to 2.0 
to be plausible. Comparison of the blue contours to the allowed
(i.e. shaded) region shows that the detected population of LAEs
can only ionize the IGM if $\log L_{*,Ly\alpha} ({\rm erg~s}^{-1}) 
< 42.4$. Following the arguments of \S~\ref{sec:uv}, the luminosity
break for the i-dropouts, $M_{*,1600}$, maps to $\log L_{*,Ly\alpha} 
= 42.60$ for $f_{Ly\alpha} = 0.5$. Most previous \lya\ studies 
have claimed $L_{*,Ly\alpha}$ greater than 42.4 (Shimasaku \et 2006;
Malhotra \& Rhoads 2004).

For a population of LAEs with $\log L_{*,Ly\alpha} ({\rm erg~s}^{-1}) 
\sgreat\ 42.5$ to ionize the IGM, galaxies much fainter than our 
detection limit must contribute to the ionizing radiation at 
$z \sim 6$.  The blue contours do not intersect the allowed region
in \fig~\ref{fig:Lmodel} for high values of $\log L_{*,Ly\alpha}$.
Extending the luminosity function to very faint limits helps
produce the stellar mass that appears to have been assembled
by $z \sim 5$ (Stark \et 2007a). The flatter the faint-end slope of 
the LAE population, the harder it becomes to ionize the IGM when 
$L_{*,Ly\alpha}$ is high. When $\log L_{*,Ly\alpha}$ 
increases much beyond 42.5 in \fig~\ref{fig:Lmodel},  the allowed
(shaded) region is intersected only by the red contours with low 
values of $\zeta$. For large values of 
$\log L_{*,Ly\alpha} ({\rm erg~s}^{-1}) \approx 43$, 
the parameter $\zeta$ must be $\sles\ 0.1$ for complete ionization,
much less if the faint-end slope is not steep.
In our simple model, when $\log L_{*,Ly\alpha}$ 
approaches 43, it becomes difficult for galaxies to ionize the IGM.  

A galaxy population described by lower values of  $L_{*,Ly\alpha}$,
however, easily ionizes the IGM. In \fig~\ref{fig:Lmodel}, 
the intersection of the blue contours with the shaded region
indicates normalizations $\log \phi_{*,Ly\alpha} ({\rm Mpc}^{-3})$ 
greater than -2.2, -2.25, or -2.37 are required for faint-end 
slopes $\alpha =  -2.0$, -1.6, or -1.2, respectively. These 
high normalization models pull the luminosity break down
to $\log L_{*,Ly\alpha} ({\rm erg~s}^{-1}) = 42.42$, 42.38, and  42.35, 
respectively.  These values of the break luminosity are below 
the detection limit of the Magellan MNS survey, so the models
produce a larger \lya luminosity density per number of galaxies
detected. If one accepts values of $\zeta$ as low as 0.1, then even
the galaxy population detected by our Magellan MNS survey 
can keep the IGM ionized at $z =5.7$. 

The volume of the Magellan MNS survey is not large enough to 
constrain $L_*$ well. One could as easily fit our points in 
\fig~\ref{fig:cumlf} with a power law instead of a Schechter 
function.  To our surprise, however, we found that the other 
LAE surveys do not appear to constrain $L_*$ all that well. 
At redshift 5.7, a straight line can be drawn through the 
luminosity function data and error bars in Figure~1 of 
Malhotra \& Rhoads (2004), particularly when the number 
densities from lensing are excluded. The fitted distribution 
is better constrained in Figure~11 of Shimasaku \et, but 
considerable leeway remains due to poor constraints at the 
bright end.  We looked to lower redshift surveys for LAEs to 
determine what values of $L_*$ might be reasonable. The MUSYC 
survey at $z \sim 3$, in fact, favors significantly lower 
values, $\log L_{*,Lya\alpha} ({\rm erg~s}^{-1}) = 42.64^{+0.26}_{-0.15}$
(Gronwall \et 2007); but inspection of their Figures~10 and 11 
suggests the luminosity break is again not that well constrained 
due to the relatively shallow survey depth. It is possible that 
$L_{*,Ly\alpha}$ is lower at $z = 5.7$ than at redshift 3 since 
$M^*_{UV}$ may fade with lookback time from redshift 4 to 6
(see Bouwens \et 2007; cf. Iwata \et 2007). 

We question whether
these lower $L_{*,Ly\alpha}$ models have been ruled out by the
data at this time.  If not, then the population of LAEs detected
could contribute a high fraction of the radiation needed to
maintain the ionization of the IGM at redshift 5.7. 
If $L_{*,Ly\alpha}$ is large, then a large extrapolation   
of the luminosity function to fainter sources is required for
LAEs to account for the ionizing photons. The faintest $z=5.7$ galaxies 
recovered by the lensing surveys have unlensed luminosities of 
$\log L_{Ly\alpha} \sim
41.4$, a factor of 2.5 higher than the value of  $L_{min}$ used to
compute \fig~\ref{fig:Lmodel}. Although there is tentative evidence 
for a steep faint-end slope (Stark \et 2007b; Kashikawa \et 2006), 
it seems likely that new technology will be required to determine 
definitively whether star-forming galaxies ionize the IGM at high 
redshift.


%

%

\section{Emission-Line Detection of High-Redshift Galaxies}

We have demonstrated that multislit, narrowband spectroscopy is
a viable technique for finding LAEs at redshift 5.7. Here we compare 
the efficiency of this sky-supression technique to other strategies using 
current instrumentation. The sensitivity limits of next generation facilities
to emission-lines are then discussed.  Hydrogen \lya\  emission will
be the brightest line where the local gas kinematics allow escape 
and the IGM neutral fraction allows significant transmission.
Our discussion focuses on $\lya$ detection over the period of 
reionization from $6 < z < 30$ but also applies to nearby lines
like HeII~1640, CIV~1550, and NV~1240. 


The canonical figure-of-merit for galaxy surveys is the product
$A \Omega$, where $A$ is the collecting area of the telescope
and $\Omega$ is the solid angle of sky observed at one time.
The time required to survey a given area of sky to a particular
magnitude is then inversely proportional to $A \Omega$ for
broad-band imaging. For emission-line surveys, the quantity to 
maximize, when sky emission is the dominant source of noise,
is $A \Omega \epsilon / (\Sigma_{\lambda} \Delta \lambda \theta^2)$,
where  $\Delta \lambda$, $\theta^2$, and $\epsilon$ represent
spectral resolution, image size, and system throughput, respectively.  
For \lya observations from redshift 4 to 20 ($\lambda_o = 0.6 -  2.6\um$),
the surface brightness of the sky, $\Sigma_{\lambda}$ is dominated by the 
zodiacal background in space. Its brightness is a strong function of
distance from the ecliptic plane and was measured with COBE (Hauser \et 1998; 
Kelsall \et 1998).  Atmospheric bands of molecular lines, mainly OH 
(hydroxyl and  $O_2$), dominate the broadband background from the ground.  
The interline background at high-dispersion is much lower than the broadband 
mean, but its physical origin and intensity are not well understood. Maihara \et 
(1993) suggest it is a few times brighter than the zodiacal background, but
Bland-Hawthorne et al. (2004) argue that most of this emission is 
scattering and diffraction of the molecular lines.  Here, we will assume 
the standard interline background from Maihara, and explore how the spectral 
resolution, image quality, and system throughput inherent to various
observing techniques affects both sensitivity and survey efficiency.

Due to the high number of interlopers in many surveys for galaxies
across the reionization era, this revised figure-of-merit does not 
completely describe the potential impact of a survey.  The time required 
for redshift confirmation, not just that used to identify candidates
must be considered. In this regard, emission-line {\it sensitivity} 
is also of relevance for surveys employing continuum-break
selection.

\subsection{Current Capabilities for Detecting Faint, Emission-Line Galaxies}


Emission-line spectroscopy with large, ground-based telescopes offers 
high sensitivity to objects with low SFRs.
The HUDF observations with HST/ACS reached an exceptional broadband depth.
The luminosity function of $i$-dropouts reaches $M_{UV,AB} = -17.5$ (Bouwens
\et 2007), a star formation rate of 0.6 \msunyr.  The \lya\ emission 
from a subsample of these galaxies was detected in 30 hr integrations with Gemini 
(Stanway \et 2006).  Fainter LAEs clearly exist as shown by their 
detection with gravitational lenses (Ellis \et 2001; Santos \et 2004). 

Here we compare current capabilities of MNS, NB imaging, and tunable filters. 
Our arguments about the benefits of spectroscopy apply equally to observations
with integral field units. The small fields of view of IFUs are inefficient 
for surveys at redshifts where the neutral fraction is low.  At $z \sim 10$,
however, the only detectable LAEs may be those within the ionized bubbles
of the brightest sources. Furlanetto \& Oh (2005) estimate physical sizes
of 27 to 227~kpc. Band-limited IFUs with diameters of 6-60\asec would be
extremely effective for studying these galaxies. 
Slitless grism spectroscopy is never more sensitive  than NB imaging because each pixel 
records background from the full transmitted bandpass. It has been used effectively 
from space owing to the lower background there and the benefits of low resolution
spectra for identifying very faint galaxies (Xu \et 2007; Pirzkal \et 2007).

\subsubsection{Narrowband Imaging vs. Spectroscopy}

For any telescope and instrument for which emission-line detection is
sky-noise limited, let us compare surveys carried out with MNS to those done
with narrowband imaging. In a fixed amount of integration time $t$, 
spectroscopic observations detect fainter emission lines because the bandpass of sky 
under the line, $\Delta \lambda$, is typically much less than the width of the filter, 
$\Delta \lambda_0$.  Maximum depth results from matching the spectroscopic 
resolution to the \lya\ linewidth. For a representative width of 150\kms, 
the factor $ \Delta \lambda /  \Delta  \lambda_0$ is approximately $ 4.1 \AA 
/ 150 \AA  \approx 6^{-2}$; and the line flux reached with MNS is
\begin{equation} \label{eqn:sensitivity}
F = F_0 \sqrt{\frac{\epsilon_0}{\epsilon} \frac{\Delta \lambda}{\Delta \lambda _0} 
    \frac{\theta}{\theta_0} \frac{t_0}{t} },
\end{equation}
where quantities denoted by subscript zero represent narrowband imaging.
Our IMACS pilot survey carried out with the facility 200-l grism collected a 6.4 -- 
12.8\AA\ band of sky under each emission line, so it did not reach maximum depth.
Many emission-line sources were unresolved by our observations, setting an upper limit 
on their size of $\theta \sles\ 0\farcs75$ FWHM.  
Better image quality would reduce sky noise further for both imaging and spectroscopy. 
The factor of $\sqrt{{\theta(slit)}/{\theta_0}}$ in Eqn.~\ref{eqn:sensitivity} a
equals $\sqrt{{1\farcs5}/{\theta_0}}$ for our pilot survey.
The gain appears larger for imaging because the sky aperture for MNS is fixed by the slit
width in one spatial dimension, yielding a linear rather than a quadratic 
decreases in sky photons with seeing. For our pilot survey, we cut slits twice as 
wide as the seeing to mitigate attenuation by the aperture. Hence, we think
$\sqrt{{\theta}/{\theta_0}}$ can be tuned to a  factor  $\sim \sqrt{2}$ 
in any seeing.  The penalty in throughput, $\epsilon$, from the addition 
of the  dispersing element is minor; for the 200-l grism, the relative 
sensitivity is reduced by just $\epsilon / \epsilon_0 \sim 0.89^2$.
We conclude that had we used our telescope time for NB imaging, 
our flux limit would be a factor of 3.0 to 4.3 times higher (i.e. less sensitive). 

When the total time required to search a given volume of space for
emission-line sources is considered, the gain in sensitivity from
MNS, relative to NB imaging,  is offset by the lower solid angle 
(per pointing) of the former. With the 200-l grism, the clear aperture 
of the IMACS-MNS experiment subtends $55.3 / 656 = 8.4\% $ of the active
imaging field. In the time required to observe 12 masks, the integration 
time for direct narrowband imaging of a single field could be increased to
$t_0 = 12 t$,  increasing the line sensitivity to match that of MNS 
with IMACS.  Narrowband imaging and MNS therefore reach comparable emission-line 
sensitivity when normalized by the volume searched per unit observing time. 
It is easy to show  that this equivalence continues to hold as spectral
resolution increases.\footnote{The higher dispersion  requires a larger separation
  between slits. The increase in depth is a single pointing is cancelled
  by the increase in the number of pointings required to map the full
  aperture of an imaging survey.}
In a given amount of time, MNS reaches considerably fainter lines than NB imaging 
but selects from a smaller volume. When the objective is to optimize the
yield of the faintest sources, the MNS technique is a better choice than
NB imaging.


The development of MNS will allow more sensitive surveys
in the 8200\AA\ atmospheric window. Detector quantum efficiency,
grism blaze, and spectral resolution can all be improved for
IMACS observations, but installation of the new E2V CCDs will
provide the single largest gain in throughput. In a deep
$\sim 25 $~hr survey, we expect to reach a line luminosity
about three times fainter than the results presented here.
The equivalent star formation rate  at redshift 5.7 is just 
$0.79 f_{Ly\alpha}^{-1}$\msunyr.

The biggest shortcoming of MNS is the time required to confirm
candidates. Most of the emission-lines discovered with both
MNS and NB imaging come from foreground galaxies. Narrowband 
imaging surveys parse the LAE candidate list using broadband colors 
and line equivalent width prior to spectroscopic confirmation.
Broadband surveys of the requisite area, however, do not go
deep enough to identify the continuum counterparts of many
emission-line galaxies discovered with MNS.  Broadband depth 
will be less of a problem for differential imaging, another 
special technique under development for LAE surveys, because 
the field of view is contiguous and therefore of smaller extent
for a given survey area. A good example is the 6\farcm8 by
6\farcm8 field of DAZzLE (Dark Ages {\it redshift} \lya\ Explorer,
Horton \et 2004) which subtends an area similar to that of our 
Venetian blinds mask and will be used to search for LAEs at
at $z \sim 8 - 10$.

Possible solutions for rapid identification of LAEs from
MNS data include resolved line profiles and equivalent widths.
Recognizing the red skew of high-redshift \lya\ emission lines
requires fairly high spectral resolution. The DEEP2 redshift 
survey provides a good example. Spectra were mostly taken with 
a 1200 line mm$^{-1}$ grating.  The observations targetted 
$z \sim 1$  galaxies with multislit masks, but the ``sky'' regions of the 
slitlets subtend an area similar to our pilot survey. Sawicki \et
(2008) demonstrate that LAEs can be culled from the serendipitous 
detections of emission-line galaxies in the ``sky'' slits based 
on the shape of the line profile {\it and} the absence of additional 
emission lines. This second criterion is eliminated for MNS, and
measurements of line asymmetry may become ambiguous at low 
signal-to-noise ratio.  Deeper broadband imaging in even a single
 band longward of the emission line appears to be the fastest way 
to parse the emission-line candidates, as emission equivalent widths
will normally be very high for \lya\ due to the combination of
the $(1 + z)$ boost and the potentially young, metal-poor 
stellar population. 

\subsubsection{Tunable Filters}

Imaging observations with tunable filters can provide the same
sensitivity and survey volume as MNS. The sky noise is comparable 
when the filter is cut down to the spectral resolution element 
of the MNS survey.  The survey volumes will also be identical. While
MNS slices the field spatially, it observes the entire redshift
depth of the atmospheric window at once.  Tunable filters image
the whole field at once -- one redshift slice at a time. The
spatial and spectral slices are the same fraction of the volume
for observations with the same camera and spectral resolution.
The advantage of imaging is exact positional information for
the emission-line sources.  The position -- wavelength degeneracy
inherent to MNS will present ambiguities when sources need
to be identified in deep continuum images.

We have taken data in the COSMOS field with the recently
commissioned Maryland-Magellan Tunable Filter (MMTF) on IMACS.
  The highest  resolution in the 8200\AA\ is about 12\AA,
similar to our IMACS-MNS survey.  We found a slightly
higher throughput with the grism than the etalon, all other 
optical elements are the same.  At redshift 5.7 both search 
techniques are viable for future searches, with the MMTF
offering better positional information and MNS offering
factor of 1.5-2.0 better line sensitivity.
Further to the red, the OH-free windows are narrower; and
tunable filters offer a substantial advantage over MNS.
Although the MMTF is not sensitive beyond the 9000 -- 9250\AA\
atmospheric window, other Fabry Perot spectrographs will be. 
The Flamingos-2 Tandem TF (F2T2) is an infrared-optimized scanning
TF, which will be mounted inside the Flamingos-2 imaging 
spectrograph of Gemini south. The spectral resolution of
$\Delta V \sim 375$\kms will deliver broader lines than 
optimal but will likely offer an efficient ground-based
survey strategy at $z > 7$ in the near future.

%


\subsection{Relation to Next-Generation Facilities}

Spectroscopy with the next-generation of large 
ground-based telescopes will reach extraordinary sensitivity
for emission-lines in the airglow band gaps. The gain from
diffraction-limited images will be as important as the increased 
aperture. Relative to our our IMACS MNS survey (subscript ``P'' for
pilot), MNS observations with  a next-generation ground based telescope 
of aperture $D$ could reach
\begin{eqnarray}
 F_{30m} = F_P / {\textsc C_{Strehl}} \sqrt{\frac{A_P}{A_{30m}} 
\frac{\theta^2_{30m}} {\theta^2_P} } \nonumber \\
 \approx (D_P/30)^2 F_P,
\end{eqnarray}
where ${\textsc C_{Strehl}}$  refers to the fraction
of total light from an object in a circular aperature of diameter 
$\theta$ and is related to the Strehl ratio.\footnote{The Strehl
ratio refers specifically to the ratio of peak intensity from a point source
to the maximum intensity in a diffraction limited image with the same total
flux.}

With adaptive optics and a 30~m aperture, line fluxes 21 times fainter 
than our IMACS observations become detectable. This sensitivity
reaches  SFRs of $0.11 f_{Lya}^{-1}$\msunyr\ at redshift 5.7 in 
10 hours.  Such a population may include the  massive, young star clusters 
whose relics we see as the globular cluster population of
the Milky Way today.

Next generation surveys may actually detect fainter line fluxes 
from higher redshift objects. First, the image quality delivered by adaptive 
optics will be better at longer wavelengths.  Second, 
at high spectral resolution, the inter-hydroxyl-line background
falls slowly with increasing wavelength between the 8200\AA\
bandpass and 1.9$\mu$ m.\footnote{The Gemini ITC background model
     suggest the between-the-lines background in units of $\Sigma_{\nu}$ 
     reaches a maximum in J band (around 1.25\um) which is 
     is about three times the 8200\AA\ background. Converting from
     frequency units to wavelength units cancels this out,
     $\frac{\Sigma_{\lambda}(1.25)}{\Sigma_{\lambda}(0.82)} = 0.43 
     \frac{\Sigma_{\nu}(1.25)}{\Sigma_{\nu}(0.82)}$.
}
Ground-based line searches should become extremely sensitive 
to emission lines from redshift 5.7 to 15.  Longward of 2\um, 
thermal emission causes the background to rise and
observations with JWST will have a strong advantage.

The ionized regions at redshift 7 to 10 will be mapped out
by pointing these large telescopes at bright LAEs and finding
fainter LAEs in the ionized bubbles.
The \lya\ line will only be transmitted 
through the IGM when the source is surrounded by an ionized
 bubble (Furlanetto \& Oh 2005).
An imaging tunable-filter, although not presently planned for 
first-light instrumentation on a 30m, could be the instrument
of choice for such surveys. A more popular option for a
facility instrument may be sets of deployable integral field
units. A combined field-of-view exceeding $ 6\asec \times
 6\asec$ could map the HII regions  and provide
spectral resolution or coverage for flagging interlopers.

The extreme sensitivities attainable at the diffraction limit
will only be realized if some \lya emission regions are small,
$\sles\ 50$~pc. Some theoretical models do predict small sizes 
(Le Delliou \et 2006), but \lya emission halos could also be quite 
extended. Empirical scaling relations for nearby HII 
regions, $R_{HII} \propto \sigma^2$ (Terlevich \& Melnick 1981),  
suggest we should expect  narrow lines of width $\sles\ 40$\kms
FWHM for small LAEs near the detection limit.  The progenitors
of the Milky Way's globular cluster systems could plausibly be 
detected as young star clusters.

\subsubsection{Ground-Based Emission-Line Sensitivity vs. JWST}

Observations from the James Webb Space Telescope will
not be hindered by telluric-OH emission, which is produced in 
a layer of atmosphere at roughly 87~km, or absorption by molecules
in earth's atmosphere, which precludes observations over about 20\%
of the 1-5\um region.  The background from space is
dominated by sunlight scattered off zodiacal dust and is strongly
dependent on ecliptic latitude and longitude. Average values are
about 10\% as bright as the Maihara (1993) model for the interline 
background from the ground, although ``deep fields'' will presumably
be carried out in regions of minimum zodi.

The JWST instruments FGS-TFI and NIRSpec provide the highest sensitivity
to line emission.  The planned configurations of these instruments do not
minimize sky background. A single MEMS aperture on NIRSpec is much larger 
than a point source, and  the spectral resolution of the Fabry Perot 
will not resolve line emission from galaxies.   These designs reflect 
the dominance of detector noise over shot noise from the sky for many 
observations. Because JWST will be cooled to $< 50$~K, emission-line 
observations with JWST will have their biggest advantage over ground-based
line surveys at $\lambda > 2$\um.  The background (in
frequency units) should reach a minimum between 3 and 4\um\ due to the
slow decline of the zodiacal background with wavelength.  In addition to
extending studies of LAEs to redshift 30, the redder (and line-free) 
spectral coverage will be critical for studying hydrogen Balmer emission 
from objects toward the end of Reionization.

Emission-line observations at $z < 15$ with JWST will not
be restricted to OH-free bandpasses but are unlikely to be as
sensitive as ground-based spectroscopy with a 30~m class telescope.
Using the background-limited case for scaling, the limiting flux from 
space relative to a 30~m telescope is
\begin{equation}
F_{JWST} = F_{30m} {\textsc C_{Strehl}} \sqrt{\frac{\Sigma_{JW}}{\Sigma_{30}} 
\frac{A_{30}}{A_{JW}} \frac{\Delta \lambda_{JW}}{\Delta \lambda_{30}} 
\left( \frac{\theta_{JW}}{\theta_{30}} \right)^2}.
\end{equation}
Assuming a relative sky brightness, $\Sigma_{30} / 
\Sigma_{JW} \sim 10$, ground-based spectroscopy could detect
lines about a factor of five fainter than JWST if the diffraction
limit of a 30~m telescope can be reached. The line sensitivity of
these two telescopes is comparble at equivalent spectral resolution
and image quality; the extra collecting area roughly canceling the
brighter sky.  Even in the JWST era, large ground-based telescopes 
may lead the search to map the progression of reionization with
LAEs.

\section{Summary} \label{sec:summary}

This paper presents the first LAEs discovered with narrowband,
multislit spectroscopy.  The technique succeeds on the 6.5~m
Baade telescope because of the large detector on IMACS and 
its wide field-of-view. Our configuration reduced the sky
noise by a factor of 10 relative to imaging through the same
narrowband filter and provided spatial multiplexing 100 times 
that of a single longslit observation. The faintest confirmed 
LAE in the pilot survey has an observed line flux of $F_{Ly\alpha} 
= 1.4 \pm 0.2 \times 10^{-17}$\flux, which is near the estimated
90\% completeness limit. 

The three redshift 5.7 LAEs have luminosities $L_{Ly\alpha} \ge\ 5 
\times 10^{42}$~erg~s$^{-1}$, which correspond to star formation
rates greater than 5-7\msunyr (for Salpeter IMF from 0.1 to 100\msun\
and solar metallicity).  The fraction of intrinsic \lya\ emission 
absorbed by the galactic ISM  likely varies with age and wind 
strength, while intervening IGM will absorb the \lya\ photons
emitted blueward of the line center, contributing to the red
skew of the line profile.  The true star formation
rates could easily be closer to 10-30\msunyr.  
 Winds with speeds of a few hundred \kms are common among local 
galaxies with star formation rates comparable to these LAEs 
and are consistent with the red wing of \lya line profile.
The rest-frame
ultraviolet continuum of these LAEs is fainter than the 
sensitivity limits of current wide-angle surveys. Future
broadband measurements will constrain the age, mass,  and 
dust content of these galaxies.  Their relation to the
i-dropout population is likely complex, although our best
estimate of the cumulative \lya\ luminosity distribution
is consistent with being drawn from the i-dropout
population around redshift 6.

Results from the IMACS MNS survey tightly constrain
the product of the luminosity function parameters 
$L_{*,Ly\alpha}$ and $\phi_{*,Ly\alpha}$. Nearly all
candidates were followed up with both low- and high-
resolution spectroscopy. The LAE population detected 
to date is unlikely to supply enough Lyman continuum 
photons to maintain the ionization of the IGM at redshift 
5.7. For this detected population of LAEs to ionize
the IGM,  both a low value of the break luminosity (
and therefore a high normalization)  
and a combination of clumping, attenuation, 
and escape factors yielding  $\zeta \approx 0.1$, where
$\zeta \equiv C_{6} (1 - 0.1 f_{LyC,0.1}) f_{Lya\alpha,0.5} 
f_{LyC,0.1}^{-1}$,  would have to be realized. Determining
whether galaxies ionize the IGM will clearly require
deeper surveys that pin down the lumnosity function
parameters.

Deeper emission-line surveys in the OH-free bandpass at 8200\AA\ 
are possible with either MNS or the MMTF at Magellan. The former 
approach offers maximum depth while the tunable filter offers
more accurate source positions. Both methods probe the same
volume per pointing, which is much larger than the volume strongly 
lensed by a cluster. These techniques can be used on
Magellan to reach lines 2-4 times fainter than the results 
presented here. Such depth over a large volume should remove 
ambiguity about the value of $L_{*,Ly\alpha}$, particularly
if complemented by shallower surveys with the volume to
count LAEs brighter than $\log L_{Ly\alpha} ({\rm erg~s}^{-1}) = 43$.  

Emission-line sensitivities will likely need to be improved 
further to find both the sources responsible for reionization 
and those that prevent the IGM from recombining at $z \sim 6$.
Finding these sources is a key science
goal of the JWST mission. The next generation
of ground-based telescopes with primary mirrors $\sim 30$~m
in diameter (and adaptive optics) will offer even
higher sensitivity to line emission shortward of $2$\um,
extending LAE searches to redshift 15 and detecting
other lines like HeII 1640 at lower redshift.  Our experiences
with MNS indicate that isolated lines, lines with very
large observed equivalent widths, and lines with red-skewed, 
asymmetric profiles are not exclusively \lya. Culling 
or exploiting the large number of foreground emission-line 
galaxies from samples will have to be a key part of any 
experimental design. While identification of the Lyman
break is one robust method, it obviates one perk of
line searches, which is discovering objects below the detection 
limit of continuum surveys.

\acknowledgments
We thank Peter Capak for discussions of the COSMOS observations,
Bahram Mobasher for his photometric-redshift catalog, Hsiao-Wen Chen
for computing limiting sensitivities of the LCIRS images, and
Richard Ellis, Rychard Bouwens, and an anonymous refereee for
helpful comments on the manuscript.  C.L.M. is grateful for the
lively environment of the Kavli Institute for Theoretical Physics,
where the manuscript was completed. This work was supported in part
by the David and Lucile Packard Foundation, the Alfred P. Sloan Foundation,
and the National Science Foundation under Grant No. PHY05-51164.


{\it Facilities:} \facility{Magellan}, \facility{Keck}


\clearpage

\begin{table}
\caption{Magellan Multislit, Narrowband Emission-Line Survey}
\begin{tabular}{lllllll}
\hline
Date &  Mask & RA & DEC & Slit PA & Exposure & Clouds/\\
     &       &  (J2000)          & (J2000)           & (\deg)     & (hr) & Image Quality \\
\hline
\hline
2004 April & COSMOS-0   & 10:00:42.9 & +02:11:00 & 90 & 6.33 & Clear; 0.8-2.5\arcsec\\
2004 April & 15H Field-0 & 15:23:35.48 & -00:08:00.00 & 135 & 10.8 & Clear; 0.8-2.5\arcsec \\
2005 March & COSMOS-0.5 &10:00:42.9  & +02:11:00 & 90 & 10.0 & Clear; 0.6-0.8\arcsec \\
2005 May   & 15H Field-0.5 & 15:23:35.48 & -00:08:00.00 & 135 & 6.75 & Thin Clouds; 0.6-0.8\arcsec \\
\hline
\end{tabular}
\label{tab:observations}       \end{table}

\clearpage

\begin{table}
\caption{Candidate Yield and Spectroscopic Classification}
\begin{tabular}{llllll}
\hline
  (1)                       & (2)     & (3)            & (4)        & (5)      & (6)      \\
  Survey                    & Confirmed  & Confirmed   & Foreground & No Follow& Not      \\
                            & LAEs       & Single Line & Objects    & Up       & Recovered\\
\hline							
\hline
10H 2004 - Line Only        &  1    & 7         & 20     & 3 & 5   \\
10H 2005 - Line Only        &  0    & 2         & 27     & 1 & 3   \\
15H 2004 - Line Only        &  1    & 6         & 25     & 6 & 4   \\
15H 2005 - Line Only        &  1    & 1         & 19     & 1 & 3   \\
10H 2004 - Line + Cont.     &  0    & 0         & 18     &16 & 1   \\
10H 2005 - Line + Cont.     &  0    & 0         & 20     &12 & 4   \\
15H 2004 - Line + Cont.     &  0    & 0         & 21     &11 & 8   \\
15H 2005 - Line + Cont.     &  0    & 0         & 19     & 0 & 1   \\
\hline
\end{tabular}
\tablecomments{
Cols -- (1) Survey name and description of the continuum flux in the NB8190 band.
(2) Number of confirmed LAEs.
(3) Number of LAE candidates confirmed to be single-line sources.
(4) Number of lines in 8190 band identifed as foreground objects.
(5) Number of emission-line objects not followed up.
(6) The Number of emission-line objects not redetected. Note that exposure times of follow vary.
}
\label{tab:class} \end{table}

\clearpage

\begin{table}
\caption{LAE Candidates}
\begin{tabular}{llllll}
\hline
(1)             &  (2)          &  (3)          &   (4)               &  (5)  & (6)          \\
Name		& RA            & DEC           & F(Lya)              & $W_o$ & $\lambda_o$  \\
		& (J2000.0)	& (J2000.0)	& (\flux)             & (\AA) & (\AA)        \\
\hline
\hline
MSDM80+3  & 10:00:30.479    & 02:17:14.80	& $14\pm2 \times 10^{-18}$  & $> 244$ & 8141 \\ 
MSDM64-5        & 10:01:12.769    & 02:13:14.80	& $63\pm8 \times 10^{-18}$     & $> 1082$& 8141      \\
MSDM13-3        & 10:00:54.458    & 02:00:30.67 & $14\pm2 \times 10^{-18}$  & $>240$     & 8169 \\
MSDM50-7        & 10:01:23.675    & 02:09:45.15 & $16\pm4 \times 10^{-18}$  & $>281$     & 8122 \\
MSDM52+8        & 09:59:53.266    & 02:10:15.11	& $25 \times 10^{-18}$       & 421     & 8179     \\
MSDM30.5-8      & 10:01:29.407    & 02:04:53.60	& $26\pm8 \times 10^{-18}$     & $> 447$ & 8173  \\
MSDM17.5+8      & 10:00:08.176    & 02:01:38.46	& $43\pm9 \times 10^{-18}$     & $> 739$ & 8129  \\
MSDM17-1        & 10:00:45.109    & 02:01:30.41 & $32\pm2 \times 10^{-18}$  & $> 555$        & 8162 \\                          
 MSDM29.5+5    & 15:22:57.880    & -00:07:36.34& $18\pm2 \times 10^{-18}$ & \nodata  & 8148 \\ 
MSDM71-5        & 15:24:08.929    & -00:10:43.06& $22\pm6 \times 10^{-18}$     &  $> 61$ &   8130  \\
MSDM89-2        & 15:24:03.935    & -00:03:06.53& $14\pm5 \times 10^{-18}$     &  $> 39$ &   8200    \\
MSDM57+4        & 15:23:19.653    & -00:03:20.27& $21\pm6 \times 10^{-18}$     &  $> 58$ &   8178  \\
MSDM36+5        & 15:23:03.246    & -00:06:38.71& $17\pm8 \times 10^{-18}$     &  $> 47$ &   8135  \\
MSDM54.5-1      & 15:23:37.234    & -00:08:36.91& $ 11\pm6 \times 10^{-18}$    &  $> 31$ &   8222    \\
MSDM55-3        & 15:23:48.230    & -00:11:11.24& $4.9\pm5 \times 10^{-18}$  &  $> 14$ &   8170  \\
MSDM94-4        & 15:24:15.670    & -00:04:16.34& $9.5\pm7 \times 10^{-18}$  &  $> 26$ &   8144    \\
MSDM70+1        & 15:23:44.420    & -00:04:56.60& $9.4\pm5 \times 10^{-18}$  &  $> 26$ &   8227    \\
\hline
\end{tabular}


NOTES - 
In addition to confirmed LAEs \m80+3, \m29.5+5, and \m71-5,
we list candidates presenting a single line in our low resolution
follow-up spectra.
(col 1) --  Object Name. Format is \m$ [{\rm Slit~Number}].[{\rm 
Field~Identifier}][\pm {\rm  Slit~Segment}]$ .
(col 2, 3) -- Coordinates. See \S~\ref{sec:search} for full description of 
uncertainties. (col 4) -- Line flux measured in search image.
(col 5) -- Measured emission equivalent width in the observed frame. An upper limit
on the continuum flux density of $5.8 \times 10^{-20}$ and  $3.6 \times 
10^{-19}$~erg~s$^{-1}$~cm$^{-2}$ \AA $^{-1}$  is
assumed for the 10H Subaru $z^{'}$ imaging and the 15H LCIRS $z^{'}$ imaging, respectively. 
An entry of nodata indicates that this imaging does not cover the
position of the object.
(col 6) -- Wavelength measured in follow-up spectrum. The uncertainty
is dominated by the placement of the object in the 1\farcs5 slit. One half
the slit width corresponds to a wavelength uncertainty of about 7\farcs5.

\label{tab:properties} \end{table}

\clearpage

\begin{table}
\caption{Cumulative Luminosity Distribution of LAEs}
\begin{tabular}{lllll}
\hline
F  & L & V  & $\phi(>L)$ 95\%CL & $\phi(>L)$ 84.1\% CL \\
(\flux) & (erg s$^{-1}$) & ($h_{70}^{-3}$ Mpc$^{-3})$ & ($h_{70}^3$~Mpc$^{-3}$) & ($h_{70}^3$~Mpc$^{-3}$) \\
\hline
\hline
%
$2.2\times  10^{-17} $& $7.8\times 10^{42}$ & $4.429\times 10^{4} $ & $2.26^{+8.5}_{-2.1} \times 10^{-5}$ & $2.26^{+5.2}_{-1.9} \times 10^{-5}$ \\ 
$1.8\times  10^{-17} $& $6.4\times 10^{42}$ & $3.904\times 10^{4} $ & $4.79^{+13}_{-3.2} \times 10^{-5} $ & $4.79^{+7.9}_{-2.8} \times 10^{-5} $ \\ 
$1.4\times  10^{-17} $& $5.0\times 10^{42}$ & $2.402\times 10^{4} $ & $8.95^{+20}_{-5.1} \times 10^{-5} $ & $8.95^{+12}_{-4.4} \times 10^{-5} $ \\ 
\hline
\end{tabular}

Notes:  Col(1) -- Estimated line flux of LAE. Col(2) --  Estimated \lya luminosity.
Col(3) Survey volume within which the object could have been detected.  
Col(4,5) Cumulative number density of LAEs computed from $\phi (L \ge L_i) = \Sigma_j n_j = \Sigma_j (1/V_{m,j})$, where the sum 
is over all galaxies $j$ with $L_j \ge L_i$.  We divide the Poisson error on 1 object by the volume $V_j$ to get
the error on each $n_j = 1 / V_j$ term.  Following Gehrels (1986), the 84.1\% (95\%) confidence range for 1 detection
is 0.173 to 3.300 (0.0513 to 4.744) objects, respectively.  And the total   error is $\delta \phi = \delta(N=1) [\Sigma_j (1/V_{m,j})^2]^{0.5}$,
where $\delta(N=1)$ takes the values  +3.744 and -0.9487 objects (in column 4) and
+2.300 and -0.827 objects (in column 5).
\label{tab:cumlf} \end{table}

\clearpage

\clearpage

\clearpage

\clearpage

\clearpage

\clearpage

\clearpage

\clearpage

\clearpage

\clearpage

\clearpage

\clearpage

\end{document}